\documentclass[longauth]{aa}
\usepackage{amssymb}
\usepackage{txfonts}
\usepackage{graphicx}
\usepackage{longtable}
\usepackage{color}
\usepackage{multirow}
\usepackage{natbib}
\bibpunct{(}{)}{;}{a}{}{,}
\usepackage{pdfpages}
\usepackage{subfigure}
\usepackage{bbding}
\usepackage{lscape}
\usepackage{xcolor}
\usepackage{url}
\usepackage[para]{threeparttable}
\usepackage{xspace}
\usepackage{float}
\usepackage[colorlinks=true, linkcolor=blue, urlcolor=blue, citecolor=blue]{hyperref}
\usepackage{booktabs}

\usepackage{afterpage}

\usepackage[utf8]{inputenc}
\DeclareUnicodeCharacter{0327}{\c{}}

\usepackage[load-configurations = abbreviations]{siunitx}

\newcommand{\codetected}{\ensuremath{\mathrm{CD}^\mathrm{BAT}_\mathrm{GBM}}}

\newcommand{\Fermi}{\emph{Fermi}\xspace}
\def\Swift{\emph{Swift}\xspace}

\author {
A.~Acharyya\inst{1}  
\and F.~Aharonian\inst{2,3}  
\and C.~Arcaro\inst{39}\textsuperscript{*}  
\and H.~Ashkar\inst{4}  
\and M.~Backes\inst{5,6}  
\and V.~Barbosa~Martins\inst{7}  
\and R.~Batzofin\inst{8}  
\and Y.~Becherini\inst{9,10}  
\and D.~Berge\inst{7,11}  
\and K.~Bernl\"ohr\inst{3}
% ...
}

\institute{University of Southern Denmark, Campusvej 55, 5230 Odense M,
Denmark
\and Astronomy \& Astrophysics Section, School of Cosmic Physics, Dublin Institute for Advanced Studies, DIAS Dunsink Observatory, Dublin D15 XR2R, Ireland
\and Max-Planck-Institut für Kernphysik, P.O. Box 103980, D 69029 Heidelberg, Germany
\and Laboratoire Leprince-Ringuet, École Polytechnique, CNRS, Institut Polytechnique de Paris, F-91128 Palaiseau, France
\and University of Namibia, Department of Physics, Private Bag 13301, Windhoek 10005, Namibia
\and Centre for Space Research, North-West University, Potchefstroom 2520, South Africa
\and Deutsches Elektronen-Synchrotron DESY, Platanenallee 6, 15738 Zeuthen, Germany
\and Institut für Physik und Astronomie, Universität Potsdam, Karl-Liebknecht-Strasse 24/25, D 14476 Potsdam, Germany
\and Université Paris Cité, CNRS, Astroparticule et Cosmologie, F-75013 Paris, France
\and Linnaeus University Sweden, Universitetsplatsen 1, 352 52 Växjö, Sweden
\and Institut für Physik, Humboldt-Universität zu Berlin, Newtonstr. 15, D 12489 Berlin, Germany
\and LUX, Observatoire de Paris, Université PSL, CNRS, Sorbonne Université, 5 Pl. Jules Janssen, 92190 Meudon, France
\and Sorbonne Université, CNRS/IN2P3, Laboratoire de Physique Nucléaire, et de Hautes Energies, LPNHE, 4 place Jussieu, 75005 Paris, France
\and IRFU, CEA, Université Paris-Saclay, F-91191 Gif-sur-Yvette, France
\and Friedrich-Alexander-Universität Erlangen-Nürnberg, Erlangen Centre for Astroparticle Physics,  Nikolaus-Fiebiger-Str. 2, 91058 Erlangen, Germany
\and Instytut Fizyki Ja̧drowej PAN, ul. Radzikowskiego 152, ul. Radzikowskiego 152, 31-342 Kraków, Poland
\and School of Physics, University of the Witwatersrand, 1 Jan Smuts Avenue, Braamfontein, Johannesburg, 2050, South Africa
\and School of Physical Sciences and Centre for Astrophysics \& Relativity, Dublin City University, Glasnevin, Dublin D09 W6Y4, Ireland
\and University of Oxford, Department of Physics, Denys Wilkinson Building, Keble Road, Oxford OX1 3RH, UK, UK
\and Aix Marseille Université, CNRS/IN2P3, CPPM, Marseille, France
\and Laboratoire Univers et Particules de Montpellier, Université Montpellier, CNRS/IN2P3, CC 72, Place Eugène Bataillon, F-34095 Montpellier Cedex 5, France
\and School of Science, Western Sydney University, Locked Bag 1797, Penrith South DC, NSW 2751, Australia
\and Université Bordeaux, CNRS, LP2I Bordeaux, UMR 5797, F-33170 Gradignan, France
\and Institut für Astronomie und Astrophysik, Universität Tübingen, Sand 1, D 72076 Tübingen, Germany
\and Universität Innsbruck, Institut für Astro- und Teilchenphysik, Technikerstraße 25, 6020 Innsbruck, Austria
\and Universität Hamburg, Institut für Experimentalphysik, Luruper Chaussee 149, D 22761 Hamburg, Germany
\and Obserwatorium Astronomiczne, Uniwersytet Jagielloński, ul. Orla 171, 30-244 Kraków, Poland
\and Landessternwarte, Universität Heidelberg, Königstuhl, D 69117 Heidelberg, Germany
\and Institute of Astronomy, Faculty of Physics, Astronomy and Informatics, Nicolaus Copernicus University, Grudziadzka 5, 87-100 Torun, Poland
\and Nicolaus Copernicus Astronomical Center, Polish Academy of Sciences, ul. Bartycka 18, 00-716 Warsaw, Poland
\and School of Physical Sciences, University of Adelaide, Adelaide 5005,
Australia, 
\and University of Leicester, School of Physics and Astronomy, University Road, Leicester, LE1 7RH, United Kingdom
\and Yerevan Physics Institute, 2 Alikhanian Brothers St., 0036 Yerevan, Armenia
\and Department of Physics, Konan University, 8-9-1 Okamoto, Higashinada, Kobe, Hyogo 658-8501, Japan
\and Kapteyn Astronomical Institute, University of Groningen, Landleven 12, 9747 AD Groningen, The Netherlands
\and GRAPPA, Anton Pannekoek Institute for Astronomy, University of Amsterdam, Science Park 904, 1098 XH Amsterdam, The Netherlands
\and Laboratoire d'Annecy De Physique Des Particules (LAPP), CNRS, 9 Chem. de Bellevue, 74940 Annecy, France 
\and Scuola Internazionale Superiore di Studi Avanzati (SISSA), Via Bonomea 265, I-34136 Trieste, Italy
\and INFN Sezione di Padova, Via Marzolo 8, 35131 Padova, Italy}

\begin{document}

\title{The second H.E.S.S. gamma-ray burst catalogue: 15 years of observations with the H.E.S.S. telescopes}

\date{Accepted XXX. Received YYY; in original form ZZZ}

\abstract 
{Recent observational efforts using imaging atmospheric Cherenkov telescopes (IACTs) have led to firm detections of very-high-energy (VHE) signals from bright gamma-ray bursts (GRBs), often at moderate redshifts.}
{This work presents 15 years of H.E.S.S. GRB observations and examines their implications through population comparisons and selected modelling cases.} 
{GRBs are a key science target of the High Energy Stereoscopic System (H.E.S.S.). With a low-energy threshold ($\lesssim$100 GeV) and rapid repointing capabilities, H.E.S.S. can begin follow-up observations within tens of seconds after a GRB trigger, covering the late prompt or early afterglow phases.}
{We report GRB follow-up observations with H.E.S.S. from 2004 to 2019, which resulted in no significant VHE signals (aside from the detections of GRB~180720B and GRB~190829A). The resulting upper limits comprise the largest set available for GRBs at VHE.}
{
A subset of bursts with favourable conditions were selected for X-ray analysis and emission modelling. Population studies were performed to compare detected and non-detected GRBs. The results indicate that VHE-detected GRBs are not a distinct population, but tend to feature luminous X-ray emission and favourable redshift and observing conditions. This highlights the potential of next-generation IACTs such as the Cherenkov Telescope Array Observatory (CTAO), whose lower energy threshold will enhance the detection of fainter and more distant GRBs. }


\maketitle

\section{Introduction}
\label{sec:intro}

Gamma-ray bursts (GRBs) are observed as brief and intense pulses of sub-MeV $\gamma$-rays, known as prompt emission, releasing up to $10^{51}-10^{54}$\,erg of isotropic equivalent energy. They are typically followed by a longer lived, slowly evolving broadband afterglow emission across the electromagnetic spectrum.
The duration of these transient events spans from milliseconds up to hundreds of seconds and GRBs are currently detected at an average rate of $\lesssim 1$ per day. 

Since their discovery in 1969~\citep{Kleb73}, GRBs have been the focus of numerous observational studies across all wavelengths, as they are excellent laboratories for studying particle acceleration at relativistic shocks.  For a long time, the very-high-energy (VHE; $E>100$\,GeV) signal associated with GRBs posed a significant challenge to imaging atmospheric Cherenkov telescopes (IACTs) from both a technical and a scientific point of view (see e.g.~\citealt{HESS_GRB_cat_2009}). 
The unpredictable occurrence of GRBs makes it difficult for IACTs to both point and start follow-up observations of these sources rapidly enough to catch their early emission phase. Detecting a VHE gamma-ray signal from GRBs has long been considered crucial for understanding the physics of these objects, particularly during the so-called early afterglow phase when the co-existence of forward and reverse shocks in the ejected outflow has the potential to yield a large variety of emission scenarios~\citep{Zhang_2018}. GRBs at VHE are also essential to understand the production of ultra-high energy cosmic rays~\citep{2015APh....62...66B}, test the extra-galactic background light (EBL) attenuation~\citep{Desai_2017} and physics beyond the standard model~\citep{PhysRevD.108.123023}, and other topics of study.

According to the widely accepted relativistic shock model~\citep[see e.g.][]{Pacz86,Piran99}, GRB emission arises from the conversion of the kinetic energy of a relativistic outflow into electromagnetic emission. Although the details of this conversion remain poorly understood, a widely accepted scenario is that the observed photons are emitted by particles accelerated at shocks internal to the relativistic outflow.  Within this framework, synchrotron emission has primarily been considered the most natural to explain the GRB sub-MeV emission~\citep[see e.g.][]{2001ApJ...548..787S,ZM01,GZ07}. 

Recent detections of significant VHE emission from GRBs were reported by the High Energy Stereoscopic System telescopes (H.E.S.S.), which observed the afterglow of GRB~180720B \citep{HESS_GRB180720B} and GRB~190829A \citep{HESS_GRB190829A}, by the Major Atmospheric Gamma Imaging Cherenkov (MAGIC) telescopes detecting GRB~190114C \citep{MAGIC_GRB190114C_1} and GRB~201216C \citep{MAGIC_GRB201216C}, as well as by the Large High Altitude Air Shower Observatory (LHAASO) detecting GRB~221009A \citep{LHAASOGRB}. These represent a significant, long-awaited result for the VHE astrophysics community and an important step forward in the understanding of GRB physics. GRB~221009A was observable by LHAASO at the burst's onset, but VHE emission was only detected 230\,s from the trigger time of the Gamma-ray Burst Monitor (GBM) on board \textit{Fermi}. In contrast, the other detections were all achieved in the early-to-deep afterglow GRB phase.

In the recent detection of GRBs at VHE, similar temporal profiles were found in the TeV, GeV, and X-ray components. Consequently, one interpretation describes this VHE emission through a synchrotron Self-Compton (SSC) mechanism. In this scenario, the synchrotron photons observed in the X-ray band and generated by a population of electrons are Compton up-scattered to MeV and GeV energies and subsequently Lorentz-boosted to GeV to TeV energies~\citep[see e.g.][]{2001ApJ...548..787S,GZ07}. Another model that describes the VHE emission in GRBs extends the synchrotron scenario up to the highest energies. This scenario can naturally explain the sub-MeV emission in GRBs, given the similar X-ray and gamma-ray flux levels seen in the VHE-detected GRBs. Still, such an extension to higher energies faces the synchrotron burn-off limit. The synchrotron burn-off limit~\citep{10.1093/mnras/205.3.593} is the maximum photon energy that electrons can radiate via synchrotron emission before their cooling timescale becomes shorter than their shortest possible acceleration timescale, preventing them from reaching higher energies. This limit is $E_{\rm max}\approx 100\Gamma$\,MeV (given in the observer's frame), where $\Gamma$ is the bulk Lorentz factor of the emission zone, which depends on its Doppler factor. In the early afterglow phase, predictions for the bulk Lorentz factor $\Gamma$ are typically of a few hundred. Therefore, explaining the GeV photons detected at late times (e.g. 10\,hours after the onset of the burst in the case of GRB~180720B) poses a challenge to pure synchrotron emission models. In this case, a Lorentz factor of $\Gamma \sim O(1000)$ would be required, whereas values of $\Gamma < 10$ are expected at such late times~\citep{HESS_GRB180720B}. This strict requirement on $\Gamma$ no longer holds if we relax the one-zone assumption and allow for multiple emission regions or extended acceleration zones~\citep{Khangulyan_2023}.
 With this alternative in mind, it has become possible to model the striking similarities between the X-ray and VHE emission seen in GRB~190829A with a pure synchrotron model~\citep[][]{HESS_GRB190829A}. 

The VHE detections of GRBs have opened a new spectral window in GRB studies, motivating further investigations of and refinements to existing models. Several past H.E.S.S. studies have reported upper limits on GRB emission at VHE~\citep{HESS_GRB_cat_2009, HESS_GRBUL_2012}, providing important constraints on emission models despite the lack of significant detections.
The observation of a large number of events, even with no detection, is thus crucial in exploring the physical parameter space of the GRB sample to understand whether all GRBs have a distinct VHE component. Observations can also help determine whether peculiar events like GRB~190829A, which is the sole low-luminosity GRB in the set of VHE-detected GRBs, belong to a distinct GRB population or whether the parameter space of possible VHE GRBs is much larger than assumed in the past. This paper presents the results of GRB observations conducted from 2004 to 2019 by the H.E.S.S. telescopes. The synchrotron+SSC emission is modelled for three selected GRBs given their bright X-ray emission and low redshift. Population studies were carried out on the sample of GRBs followed up by H.E.S.S., the population of VHE-detected GRBs and the whole sample of \Swift/BAT and \Fermi/GBM GRBs.

The paper is organised as follows. First, a description of the H.E.S.S. experiment and the GRB alert system is given in Sect.~\ref{sec:follow-up}. The selection method and properties of the observed GRBs are presented in Sect.~\ref{sec:observed}. Details on the data analysis are described in Sect.~\ref{sec:analysis}. The results of the data analysis are presented in Sect.~\ref{sec:results} and placed in context with the population of \Swift- and \Fermi-detected GRBs, finding three GRBs in our sample with highly constraining VHE upper limits (ULs). A one-zone SSC modelling is performed on these three GRBs using the \Swift/XRT data and H.E.S.S. ULs. An analysis of the X-ray characteristics of these GRBs in context with the overall population and VHE-detected GRBs is also presented in that section. The results of these studies are discussed in Sect.~\ref{sec:discussion} and conclusions are presented in Sect.~\ref{sec:conclusions}

\section{The H.E.S.S. telescopes and the GRB follow-up programme}
\label{sec:follow-up}

\subsection{The H.E.S.S. experiment}
\label{sec:hess}

H.E.S.S.\footnote{\url{https://hess-experiment.eu/}} is an array of IACTs located in the Khomas Highland of Namibia ($23^\circ16'18,4''$~S, $\,16^\circ30'0.8''$~E, at 1800\,m altitude a.s.l). It comprises four 12-m diameter Cherenkov telescopes (CT~1-4) placed in a square configuration of 120\,m lateral length and a 28-m diameter telescope (CT~5) located at the centre of the array. This configuration was selected to maximise the sensitivity to perform observations of gamma-ray sources with energies $\geq$ 100\,GeV up to 10\,TeV~\citep{HESS_performanceCT5}. The field of view of CT~5 ($\sim3.2^\circ$) and CT~1-4 ($\sim5^\circ$), together with a fast slewing speed of $\sim 100^\circ$/min enables the telescopes to be redirected to any part of the sky in less than 2 minutes~\citep{BOLMONT201446}. Above 100\,GeV, H.E.S.S. can detect a point-like source of flux 1.4 x 10$^{-11}$\,erg\,cm$^{-2}$\,s$^{-1}$ (3.5\% of the Crab Nebula flux) at a 5$\sigma$ level in 2 hours of observation~\citep{2006A&A...457..899A}. All these characteristics make H.E.S.S. a powerful instrument for observing GRBs at VHE. H.E.S.S. is currently the only IACT array in the Southern Hemisphere with an active GRB observation programme. Its success has been proven by detecting GRB~180720B \citep{HESS_GRB180720B} and GRB~190829A \citep{HESS_GRB190829A} and following up on many other interesting transient alerts.
In September 2019, the camera of CT~5 was upgraded with FlashCam, one of the prototype cameras foreseen for the Medium-Sized Telescopes of the Cherenkov Telescope Array Observatory~\citep[CTAO,][]{Puhlhofer19}. The results presented in this work correspond to GRBs observed before installing the FlashCam in CT~5.

\subsection{Alert system and GRB follow-up}
\label{sec:followup}
GRB alerts are sent by space and ground-based facilities through the General Coordinates Network (GCN)\footnote{\url{https://gcn.nasa.gov/}}. The GCN system distributes machine-readable alerts (GCN Notices) directly to subscribed experiments such as H.E.S.S., while also making detailed human-readable circulars available online to support rapid follow-up observations of transients. Within the GRB observation programme of H.E.S.S., the transients follow-up system~\citep{Hoischen22} performs filtering on alerts by GRB-detecting satellites such as the Neil Gehrels \Swift Observatory~\citep{swiftGRB} and the \emph{Fermi} Gamma-ray Space Telescope~\citep{Atwood09}. H.E.S.S. also followed up on alerts from the High Energy Transient Explorer Mission (HETE-2) until 2008, when the mission ended operations. 
The observational follow-up of GRBs with H.E.S.S. is fully automated in case alerts are received during regular telescope operations or observable within tens of minutes of delay and manually scheduled later when the burst is not immediately observable. For all cases, one of the GRB advocates within the H.E.S.S. collaboration serves every observing period, which covers the time between two successive full moons. The function of the burst advocate is to monitor the communication channels (GCN circulars, Astronomer's telegrams, etc.) to identify additional alerts to follow up, and consider new information that might affect planned GRB observations such as updated localisations or redshift determination. The burst advocate works with the crew on site, requesting observations to ensure that each GRB alert is appropriately followed up, communicating updated GRB coordinates to the shifters and deciding whether to extend or interrupt the follow-up observations. 

For burst alerts that are immediately observable during the night, selection criteria are applied to prioritise the most promising events, balancing the limited available observation time against constraints such as dark time and competing science programmes. The criteria for GRB observations have evolved over the lifetime of H.E.S.S. The automatic repointing criterion generally requires a GRB to be observable immediately, at a zenith angle less than $60^\circ$ (i.e. elevation greater than $30^\circ$), and for at least 10 minutes. Observations are generally conducted with four runs ($\sim$28~min each), or until the target exceeds a zenith angle of $60^\circ$.
In mid-2016, a real-time analysis (RTA) system was deployed at H.E.S.S., which achieves comparable sensitivities to the off-site analysis~\citep{Hoischen22} within 25\% uncertainties.  The GRB advocate may decide to extend the standard observation duration if the RTA system shows a significant signal in the sky maps or if information reported in GCN circulars indicates observational features that justify extended follow-up, such as a measured redshift, high-energy (HE; 100\,MeV--100\,GeV) emission detection, or a bright X-ray afterglow. \\
In the case of follow-ups triggered later by the burst advocate on-call, the bursts must be visible at a zenith angle less than $45^\circ$. A time-delay-dependent redshift cut is made for these observations, accounting for the absorption of VHE gamma rays due to their interaction with the EBL. Observations of GRBs at $z\leq0.1$ may be performed up to 24 hours after the onset of the burst, with a delay of up to 12 or 6 hours for a burst at $z\le 0.3$ and $z\le 1.0$, respectively. In the case of an unknown redshift, the maximum delay is 4 hours. This set of criteria applies to the observations presented in this paper and after the detection of GRB~180720B, detected at T$_0$+10\,h, this criterion was relaxed for well-localised GRBs (\Swift and \Fermi/LAT detections), allowing for follow-up observations with only 24 hours of delay. The GRB advocate can override these criteria if multi-wavelength observations justify such a decision. \\
Standard GRB follow-ups are performed for a maximum of 2 hours of duration. 
For triggers issued by the \textit{Fermi}/GBM, a cut is applied, selecting only those bursts with a significance greater than $10\sigma$ and a localisation uncertainty below $2^\circ$ (statistical only), based on the Ground or Final Notices. These criteria are designed to reduce the number of \textit{Fermi}/GBM follow-ups while retaining alerts with a higher probability of VHE detection. Over the years, these selection criteria have been refined and sometimes becoming more restrictive.
Although automatic repointing is also done for \Fermi/GBM alerts, which have significant localisation uncertainties, the GRB advocate on-call pays special attention to GCN notices, providing more accurate coordinates and instructing the shifters to point the telescopes to an updated position.\\ 
Under the latest selection criteria explained above, an average of only 1 or 2 GRB follow-up observations were performed by H.E.S.S. per month. The observations in this paper were conducted with all telescopes available at the time of observation and carried out in wobble mode~\citep{Fomin94} with an offset of $0.5^\circ$ in right ascension and/or declination from the GRB location.

\section{The observed GRBs}
\label{sec:observed}

\subsection{The sample of H.E.S.S. GRB follow-ups}
\label{sec:grbsample}

To extend the analysis sample beyond the observations that fell under the dedicated GRB programme described in Sect.~\ref{sec:followup}, the list of all H.E.S.S. observations conducted between 2004 and 2019 was correlated spatially and temporally with existing GRB catalogues searching for potential observations by chance. 
We used the following catalogues: 1) public \Swift GRB table,\footnote{\url{https://swift.gsfc.nasa.gov/archive/grb_table/}} 
2)  \Fermi/GBM online burst catalogue,\footnote{\url{https://heasarc.gsfc.nasa.gov/W3Browse/fermi/fermigbrst.html}} 
3)  INTEGRAL catalogue,\footnote{\url{https://www.isdc.unige.ch/integral/science/grb\#ISGRI}} 
4)  MAXI catalogue,\footnote{\url{http://maxi.riken.jp/grbs/}} and 5) the HETE-2 catalog.\footnote{\url{https://space.mit.edu/HETE/Bursts/}} 
A sky separation of 2$^\circ$ and up to 48 hours of observation delay were used for the correlation. For matching the \Fermi/GBM catalogue, the strategy was slightly different due to the large localisation uncertainty. The maximum observation delay was set to 24 hours, and required to cover at least 10\% of the localisation uncertainty region with a minimum of 0.5\% per observation run. In this sample, GRB~180720B and GRB~190829A, both detected by H.E.S.S., have not been considered here as they were previously discussed in dedicated publications \citep{HESS_GRB180720B, HESS_GRB190829A}. 
With this method, two more \Fermi/GBM GRBs were added to our sample compared to the bookkeeping of GRB observations as the sky region around the location was observed by coincidence during a different science campaign of H.E.S.S.: GRB~120218276 (removed from the analysis after data quality checks, see Sect.~\ref{Sec:quality}) and GRB~160113A. 
It is worth noting that the sky region of GRB~170730B and GRB~170826B (referred to as GRB~170826819 in Table~\ref{tab:analysisGRBsGBM}) overlap with the field covered during observations of PKS~2155-304. This correlation method resulted in 107 GRBs observed in the 15 years considered. 

Several GRB follow-ups from this set of observations were removed from the data analysis presented in this paper. Either they were no longer classified as a GRB after the H.E.S.S. observation took place \citep[for example, GRB 120625 observed with H.E.S.S. was later identified as a Galactic transient,][]{GCN13386}, or the provided alert position was later updated and the source was no longer in the H.E.S.S. field of view.

\begin{center}
\begin{figure}
\centering
\includegraphics[width=0.4\textwidth]{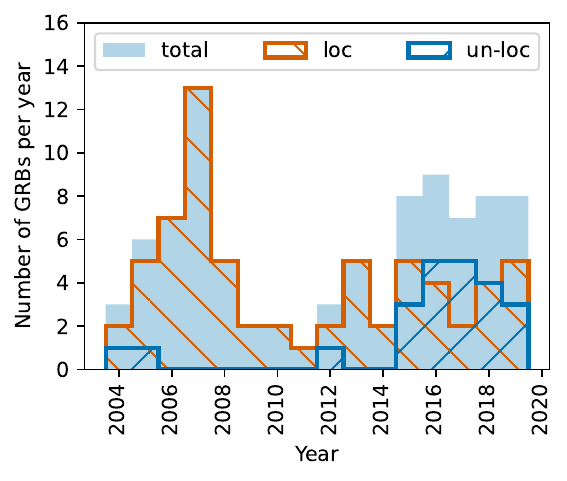}
\caption[]{Distribution of GRB follow-ups performed with H.E.S.S. between 2004 and 2019. A total of 89 GRBs are considered in the analysis, classified as either well-localised (loc, typically \textit{Swift}/BAT alerts with position uncertainty $\lesssim$3$'$) or poorly localised (un-loc, mostly \textit{Fermi}/GBM alerts with uncertainty $\gtrsim$0.2$^\circ$) as detailed in Sect.~\ref{sec:grbsample}.}
\label{fig:GRByear}

\end{figure}
\end{center}

In total, 89 GRBs were considered in the analysis and classified into two categories: The well-localised GRBs (loc), usually \textit{Swift}/BAT alerts, where the uncertainty on the position is typically 1--3 arcminutes, and thus smaller than the H.E.S.S. point-spread function (PSF, $\sim$0.2$^\circ$ for CT 5).
 The poorly localised GRBs (un-loc), mostly \Fermi/GBM alerts, comprise GRBs whose position uncertainty exceeds the PSF of H.E.S.S. This splits the sample into 66 loc and 23 un-loc GRBs. Fig.~\ref{fig:GRByear} shows the distribution of these observations over time. A noticeable spike in GRB observations by H.E.S.S. around 2006–2007 can be attributed to the sharp increase in well-localised GRB alerts following the launch of \textit{Swift} in late 2004, combined with an internal shift within the H.E.S.S. collaboration around 2007 toward more permissive follow-up criteria, which allowed observations with up to 24\,hours of delay. Another increase in follow-ups can be seen around 2013 after the commissioning of CT~5 began and again at around 2016, corresponding to when the H.E.S.S. transients follow-up system was fully commissioned, and probably a change in trigger criteria.  Table~\ref{tab:grb_population_analysis} summarises the GRB population in this work at each analysis step.

\subsection{Data quality checks}
\label{Sec:quality}

To minimise the number of trials introduced during the analysis, a detailed low-level data quality check was performed for the observation of each GRB in the sample. This allows for issues to be identified before performing the data analysis and prevents rerunning the analysis of a GRB with different analysis configurations. 
The low-level checks include inspecting trigger rates, per-camera-pixel pedestals, night-sky background levels, and centre-of-gravity maps of the reconstructed shower images. 
Based on these data quality checks, we identified runs where the telescope cameras malfunctioned due to various hardware issues or miscalibrations. In addition, these checks verified the presence of bright stars near the GRB position. Bright stars can bias the reconstruction of shower images and create spurious signals in the analysis. The location of bright stars is therefore masked during the background estimation. In addition to these checks, observations carried out during the presence of clouds in the field of view were removed from the analysis, as clouds increase the energy threshold and decrease the reliability of the spectral analyses.

For the loc sample, nine GRBs were rejected due to calibration or data-taking issues and nine due to bad weather conditions. For the un-loc GRBs, eight were rejected due to calibration issues or due to bad weather conditions (see Table~\ref{tab:grb_population_analysis}).
 
The main properties of the remaining GRBs in the sample are summarised in Table~\ref{tab:observationGRBs}. Multi-wavelength information was retrieved from the GCN Circulars repository\footnote{\url{https://gcn.nasa.gov/circulars}} and is also listed in this table. Redshift measurements and follow-ups from radio telescopes are sparse in the GRB sample of this study. This is especially the case for alerts triggered by the \Fermi/GBM, which are not often followed-up on by optical and radio facilities as their fields of view is generally much smaller than the \Fermi/GBM localisation uncertainties.

For GRBs that occurred between 2004 and 2007 and were observed by H.E.S.S., the population presented in the first H.E.S.S. GRB catalogue~\citep{HESS_GRB_cat_2009} differs from that discussed in this paper. In this work, four GRBs were removed and three newly added. Two of the four excluded GRBs fall into the period of H.E.S.S. construction, when not all four telescopes were commissioned yet (GRB~030329 and GRB~030821), while the other two (GRB~060403 and GRB~070429A) were excluded due to camera problems that were likely not identifiable with the calibration algorithms available at the time. The three GRBs now included due to different selection criteria are GRB~050607, GRB~060728, and GRB~070920B.

\begin{center}

    \end{center}

\section{Data analysis}
\label{sec:analysis}

The observation runs for each GRB are separated into different clusters to perform the analysis. If the observations started less than 10 minutes after the burst, the first run forms one cluster and is analysed separately. Observations are separated into two clusters if there are more than 4 hours between runs. This method prevents the integration of background events from late observations that might occult any potential VHE signal during the early afterglow phase. For \emph{un-loc} GRBs for which the H.E.S.S. pointing was updated during observations, the disconnected fields of view were divided into different clusters.\\
The analysis of each GRB was performed using two independent analysis chains available within the H.E.S.S. collaboration to cross-check the results. In this paper, the results from the \texttt{Model++} analysis method with the \texttt{ParisAnalysis} calibration chain~\citep{DeNaurois09} are presented, while the cross-check was obtained using the \texttt{ImPACT} analysis method with the \texttt{HAP} calibration chain~\citep{Parsons14}. These analysis methods were not available during the preparation of the first H.E.S.S. GRB catalogue, therefore, they represent an improvement in sensitivity compared to the methods used in that earlier work.

Several event-reconstruction modes are available within the H.E.S.S. analysis software: \texttt{stereo}, when performing stereoscopic reconstruction with CT 1-4 only \citep{2006A&A...457..899A}, \texttt{hybrid} when performing stereo reconstruction with all five telescopes, and \texttt{mono} when performing a reconstruction with CT~5 only~\citep{Holler15}. All three configurations were used in this analysis. For a given cluster, the choice of the reconstruction mode was determined using the following rules: For the \emph{loc} GRBs or GRBs with a localisation better than 1~degree, the \texttt{mono} analysis configuration was the preferred choice as it provides the lowest energy threshold. For the un-loc GRBs, the \texttt{hybrid} analysis was chosen for the better off-axis performance than \texttt{mono} while lowering the energy threshold compared to the analysis with the \texttt{stereo} configuration. 
Finally, for all GRB observations conducted without the use of CT~5, whether taken prior to its construction, excluded for technical reasons, or omitted from the analysis due to data quality concerns, the \texttt{stereo} reconstruction method was applied. The profile used for each analysis is described in Table~\ref{tab:analysislocGRBs}. For the gamma-ray events selection, a set of {\tt loose} cuts \citep{2006A&A...457..899A} was chosen for all analyses to lower the energy threshold at the expense of losing some energy resolution.

The analysis procedure strongly depends on whether the GRB is well-localised. In both cases, significance maps were produced using the so-called ring background method. In contrast, for the production of integral flux ULs of a loc GRB, the reflected background method was employed \citep{Berge07}. The significance of the gamma-ray emission was computed with the standard Li~\&~Ma method~\citep{Li:1983fv}. ULs were obtained at a 95\% confidence level (CL) using the \emph{Rolke} method~\citep{Rolke2005}. Integral flux upper-limits maps were produced for non-detected GRBs of the un-loc sample using a 95\% CL and assuming a power-law spectrum with photon index of $\alpha$=-2.5. ULs for the loc sample were also computed with a power-law spectrum assumption with $\alpha=-2.5$ and $\alpha=-5.0$. at 95\% CL. All ULs were obtained without applying corrections for the EBL absorption. The integral flux upper-limits were integrated from E$_{\rm thr}$ to infinity. We note that the choice of spectral index or whether to include a term accounting for EBL absorption does not affect the results of the differential flux upper limits used specifically in Sect.~\ref{sec:specificgrbs}, which are independently corrected for EBL attenuation.

\section{Results }
\label{sec:results}

The search for significant VHE emission in the GRB sample did not yield any new GRB detections. The results of this analysis are summarised in Table~\ref{tab:analysislocGRBs} for the loc sample. For the un-loc sample, the delay of observation and type of analysis configuration are summarised in Table~\ref{tab:analysisGRBsGBM}, the UL maps (figures and \emph{FITS} files) are provided in the data repository accompanying this publication\footnote{\url{https://hess-experiment.eu/publications/}}.

\begin{figure}
\centering
\includegraphics[width=0.4\textwidth]{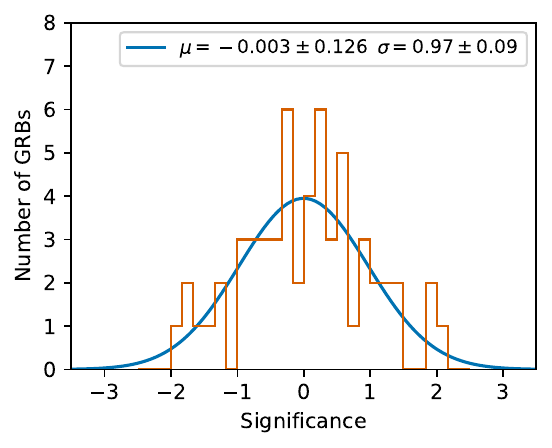}
\caption[]{Significance distribution of the gamma-ray emission for the loc follow-ups. The distribution is shown in orange, and the fitted Gaussian is shown in blue. Each entry of the histogram corresponds to one GRB or cluster (see text for details).}
\label{fig:distribsignif}
\end{figure}

Figure~\ref{fig:distribsignif} shows the distribution of the statistical significance of the VHE emission for the loc GRB sample. As no significant detections are provided, the distribution is consistent with a normal distribution centred at $\mu = -0.003 \pm 0.126$ with a standard deviation of $\sigma = 0.97 \pm 0.09$.

A stacked search for VHE emission was performed using GRBs with well-localised positions. For this, the $\theta^2$ distributions (squared angular distance between the reconstructed event direction and the GRB position) from individual GRB analyses were combined. Specifically, the ON and OFF event $\theta^2$ arrays from each GRB were summed, and a classical Li~\&~Ma significance was computed using a $\theta$ cut of $0.12^\circ$~\citep{Berge07}. This procedure effectively corresponds to stacking the total counts in a single reflected background region per observation, rather than combining multiple reflected background regions across the dataset. 

\begin{figure*}
\centering
\includegraphics[width=.9\textwidth]{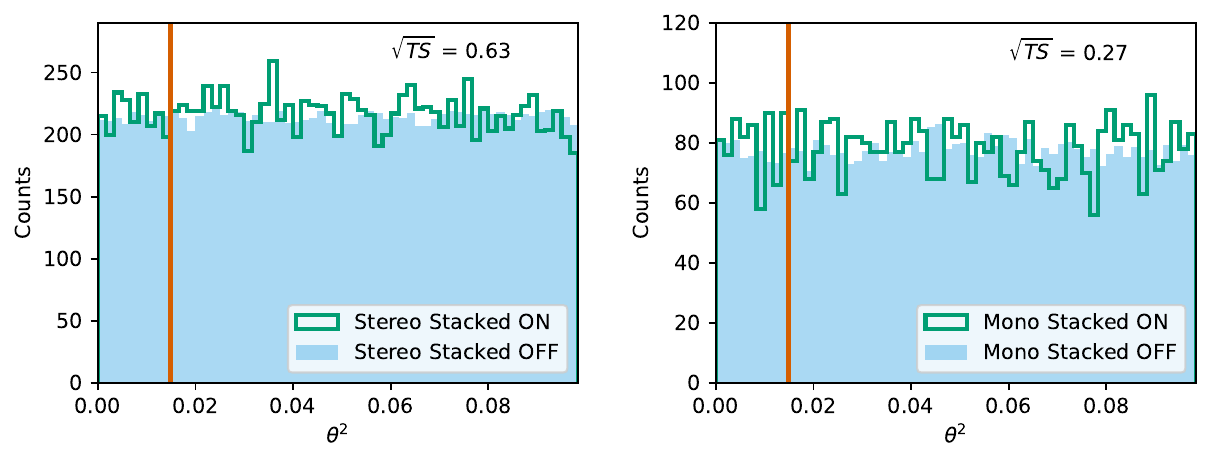}
\caption[]{ Stacked analysis of the GRB sample. The $\theta^2$ plots of stacked H.E.S.S. ON and OFF events are shown in green and shaded blue, respectively, for the stereo (left panel) and mono (right panel) observations. The orange vertical line corresponds to the $\theta^2$ cut as explained in Sect.~\ref{sec:results}.}
\label{fig:thetaclasses}
\end{figure*}

The stacked analysis was carried out separately for observations performed with CT~5 only (mono) and for those using the CT~1–4 array (stereo). The mono dataset comprises 12 GRB clusters, with a total livetime of 11.1 hours, while the stereo dataset includes 47 GRB clusters, corresponding to a livetime of 60.2 hours. In both cases, no additional cut on observation delay was applied beyond the initial selection of well-localised GRBs. The resulting $\theta^2$ distributions are shown in Fig.~\ref{fig:thetaclasses}, yielding statistical significance values of 0.63 and 0.27 for the stereo and mono datasets, respectively.

\subsection{Specific GRBs}
\label{sec:specificgrbs}

This section presents our emission modelling for a subset of GRBs selected based on how constraining the H.E.S.S. upper limits are expected to be in the context of the SSC scenario. These GRBs have good observation conditions (low zenith angle, short follow-up delay), high X-ray flux, and low redshift. Several criteria were set to select them. The first restricts the selection to only  GRBs with known redshift. The second criterion considers only GRBs with H.E.S.S. data within the first 1000\,s after T$_0$. Additionally, only GRBs whose \Swift/XRT fluence within this 1000\,s is above $10^{-9}\,\mathrm{erg}\,\mathrm{cm}^{-2}$ are selected. Defining $t_\mathrm{del}$ (in units of seconds) as the delay since the GRB trigger, all GRBs for which the X-ray flux exceeds a level of $3.2\times10^{-5} t_\mathrm{del}^{-1.2} \,\mathrm{erg}\,\mathrm{cm}^{-2} \mathrm{s}^{-1}$ were also considered. This flux threshold corresponds to the lowest flux measured by \Swift/XRT on GRB~201216C, the faintest VHE GRB, while the power-law temporal decay index is the standard index for GRB afterglows~\citep{2006ApJ...642..389N}. Only five GRBs observed by H.E.S.S fulfilled these criteria. To ensure that EBL absorption does not suppress any possible VHE emission, only GRBs for which the EBL absorption, evaluated at the energy threshold of the H.E.S.S. analysis, was lower than 90\% were kept. The EBL model from \cite{franceschini} was used to apply this selection. This final criterion yields three GRBs selected for the modelling investigations: GRB~100621A, GRB~131030A, and GRB~161001A.

Among other interesting GRBs followed up by H.E.S.S. is the long GRB~160310A (T$_{90}$ = 18.2\,s, where T$_{90}$ is the duration over which 90\% of the prompt gamma-ray fluence is detected) that was detected by both \textit{Fermi}/GBM and \textit{Fermi}/LAT, with the latter detecting a 30\,GeV photon at T$_0$ + 5800\,s. No multi-wavelength information is available for this GRB at these times.

For the three selected GRBs, the \Swift/XRT data of the time intervals matching the H.E.S.S. observation windows were retrieved with the Time Slicing tool in the XRT data repository~\citep{2009MNRAS.397.1177E}. A \Swift/XRT spectrum was extracted using the python interface of XSPEC-v12.14.0: \emph{PyXpec}. The spectrum of each GRB in the energy band of 0.3-10\,keV was fitted with the model \texttt{TBabs * zTBabs * powerlaw} using the Cash statistics (cstat). The component \texttt{TBabs} given by the model in \cite{wilms2000absorption} estimates the line-of-sight Galactic absorption, having as a parameter the Galactic hydrogen column density $N_{\rm H, gal}$, which is calculated from \cite{Willingale2013}. The component \texttt{zTBabs}  accounts for the intrinsic absorption from the GRB's local environment and its host Galaxy and depends on the redshift $z$. The fitted parameters are the intrinsic column density $N_{\rm H,int}$, power-law normalisation $\phi_{\rm XRT}$ at the reference energy $E_0=1$\,keV, and index $\alpha_{\rm XRT}$.  The H.E.S.S. ULs are computed above the energy threshold indicated in Table~\ref{tab:analysislocGRBs} and EBL-corrected with the model from \citet{franceschini}.

For the modelling, a standard single-zone synchrotron self-Compton model was adopted \citep{1998ApJ...497L..17S, 2001ApJ...548..787S}, in which a relativistic shock propagating into the external medium accelerates electrons that emit photons.
 The modelling consisted of fitting the analytical expressions of the SSC and synchrotron components of the SED, looking at explaining the XRT data with the synchrotron component. The obtained set of reference parameters were not necessarily the best-fit values, but instead chosen to represent a plausible scenario to explain the X-ray emission with the H.E.S.S. ULs. The evolution of the relativistic shock was computed following \citet{1976PhFl...19.1130B}, then the electron spectrum was obtained, and, finally, the corresponding synchrotron and SSC emissions were calculated analytically (detailed calculations can be found in \cite[e.g.][]{1998ApJ...497L..17S, 2001ApJ...548..787S, 2022ApJ...925..182H}). It is worth noting that Klein-Nishina effects were included in the calculation of the SSC component, but the feedback of Klein-Nishina suppression on the electron cooling was not accounted for; thus, possible modifications to the synchrotron spectral shape are not considered. In practice, however, the SSC contribution is found to be subdominant for the three GRBs analysed in this paper, therefore, synchrotron is the dominant cooling mechanism. 
 
 In this modelling, two scenarios were considered for the surrounding medium of the GRB: a constant interstellar medium (ISM) and a wind-like profile. The ISM environment assumes a constant-density medium. A wind-like medium follows a density profile (n= $A$/$r^{2}$), where $A$ is a normalisation factor and $r$ is the radius, characteristic of a massive star's stellar wind, leading to different afterglow evolutions. The fitted parameters are the magnetic partition ratio $\epsilon_B$ (ratio of the energy density of the magnetic field generated in the GRB shock to the total kinetic energy density of the shock), the electron partition ratio, $\epsilon_e$ (ratio of the kinetic energy density of electrons to the total energy density of the shock), and the power-law index $p$ of the non-thermal electron distribution. 
 The total kinetic energy of the shock, $E_{\rm sh}$, is restricted to a parameter range which implies that the prompt emission energy corresponds to 1\%-10\% of this energy ~\citep[see e.g.][]{Zhang_2007,10.1093/mnras/stv2033}. This assumption links the afterglow modelling to the prompt-phase energetics and ensures that the inferred values of $E_{\rm sh}$ remain within physically plausible ranges. Specifically, the modelling adopts an efficiency of 1\% for GRB 100621A, and 10\% for GRB 131030A and GRB 161001A. The magnetic ($\epsilon_B$) and electron ($\epsilon_e$) partition ratio parameters are not assumed to sum to unity, as a significant fraction of the kinetic energy is typically carried by non-radiating particles such as protons or remains in the bulk motion of the shocked material~\citep[e.g.][]{2001ApJ...548..787S}.

For simplicity, and following standard practices in GRB afterglow modelling, the injection fraction is assumed to be $\eta_{\rm inj}=100\%$, i.e, all the electrons swept by the shock contribute to the radiation spectra~\citep{Eichler_2005,1998ApJ...497L..17S}. This assumption removes an otherwise poorly constrained parameter and avoids additional degeneracies in the modelling parameters.  The resulting reference parameters of the three selected GRBs are listed in Table~\ref{tab:modelling}. This modelling considers acceleration at the Bohm limit, which is required to impose the synchrotron burn-off limit. 

\begin{table*}
    \centering
        \caption[]{Modelling parameters definition and results for the three selected GRBs} \label{tab:modelling}
    \begin{tabular}{lccc}
    \toprule
         
         Parameter & GRB100621A & GRB 131030A & GRB161001A \\
         \midrule
         Observation delay t (s)$^{*}$ & 1500 & 1070 & 375 \\
        
        Injected electron-spectrum index $p$ & 2.84 & 2.54 & 2.5\\

        Explosion shock energy E$_{\rm{sh}}$ (\,erg) & $3\times10^{54}$ &  $3\times10^{54}$ & $2\times10^{53}$ \\
         
        \midrule
         \multicolumn{4}{c}{ISM case}\\
        \midrule

         Magnetic partition fraction $\epsilon_B$ & 3 $\times 10^{-4}$  & 1.5$\times 10^{-3}$ & $5\times10^{-3}$ \\
         Electron partition fraction $\epsilon_e$ & 0.025 & 0.04 & 0.045 \\
         Number density of ambient medium $n_{0}$\,($\mathrm{cm^{-3}}$) & 0.01  & $1\times 10^{-3}$ & $1\times 10^{-3}$  \\
         \midrule
                  \multicolumn{4}{c}{Wind case}\\
        \midrule

         Magnetic partition fraction $\epsilon_B$ &0.01  & 0.02 & 0.035 \\
         Electron partition fraction $\epsilon_e$ &0.016 &0.045 & 0.05 \\  
         A ($ \rm cm^{-1}$)  & $3\times 10^{33}$ & $1\times 10^{33}$  & $3\times 10^{32}$ \\
    \bottomrule 
    
    \end{tabular}  

\tablefoot{These definitions follow the convention of~\cite{1998ApJ...497L..17S, 2001ApJ...548..787S, 2022ApJ...925..182H}. In all cases, the injection fraction $\eta_{\rm inj}$ is 100.0\%. Parameters followed by a $^{*}$ are set fixed during the analytical fitting.}
\end{table*}

\subsubsection{GRB~100621A}

\begin{figure}[H]
    \centering
    \includegraphics[width = 0.45\textwidth]{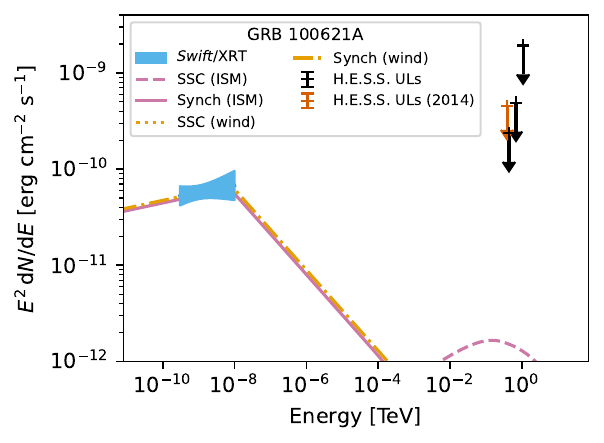}
    \caption{SED of GRB~100621A. The \Swift/XRT spectrum is shown with the blue butterfly. The H.E.S.S. ULs obtained in this work are shown in black. The H.E.S.S. UL published in \cite{2014A&A...565A..16H} is shown in orange for comparison. For the modelled emission components, the synchrotron is shown with a pink solid line (ISM case) and yellow dashed-dotted line (wind case), and the SSC with a pink dashed line (ISM case) and yellow dotted line (wind case). }
    \label{img:grb100621A}
\end{figure}

GRB~100621A is a remarkable burst, featuring an extremely bright X-ray afterglow, which was, at the time, the brightest X-ray transient ever detected by the \textit{Swift}/XRT.
 The \Swift/BAT detected this source on June 21, 2010, at 03:03:32 UTC (T$_0$) \citep{2010GCN.10873....1E}. 
 The T$_{90}$ is (63.6 $\pm$ 1.7) s and E$_\mathrm{iso}$ 
 is 2.8$\times10^{52}$ erg (20\,keV--2\,MeV) ~\citep{2010GCN.10882....1G}. Using the Very Large Telescope (VLT) equipped with the X-shooter spectrograph, the redshift was determined to be $z = 0.542$~\citep{2010GCN.10876....1M}.  
The gamma-ray spectrometer satellite Konus-Wind~\citep{KonusWP} measured a fluence up to T$_0$+26.9\,s of (3.6$\pm$0.4) $\times 10^{-5}$ erg\,cm$^-2$ in the energy range of 20 keV - 2~MeV~\citep{2010GCN.10882....1G}. The \Fermi spacecraft could not observe the burst due to occultation by the Earth. The H.E.S.S. observations began at 03:14:55 UTC, 683 seconds after T$_0$. Due to the moonrise, only two observation runs were taken.
For the XRT spectral fit, an N$_{\rm H,gal}$ value of 3.2$\times 10^{20}\,\mathrm{cm}^{-2}$ was used; the fitted intrinsic column density results in a value of N$_{\rm H,int}=(2.69\pm0.25)\times10^{22}\,\mathrm{cm}^{-2}$. The parameters of the power-law fit are $\alpha_{\rm XRT} = 1.92\pm 0.135$, $\Phi_{\rm XRT} = (3.58 \pm 0.587)\times10^{-2}$\,keV$^{-1}$\,cm$^{-2}$\,s$^{-1}$. The emission modelling results are listed in Table~\ref{tab:modelling}.
In Fig.~\ref{img:grb100621A}, we show the SED of this GRB with the H.E.S.S. upper limits and the \Swift/XRT spectrum. Previous upper limits on the energy output reported by \citet{2014A&A...565A..16H} are also shown for comparison.

\subsubsection{GRB~131030A}

The \Swift/BAT initially identified GRB~131030A on October 30, 2013, at 20:56:18 UTC (T$_0$)~\citep{GCN131030A}. The Konus-Wind instrument detected this GRB up to an energy of 10~MeV. The light curve from $~$T$_0$-3 s to $~$T$_0$+25~s showed a multi-peaked pulse, with a fluence of (6.6 $\pm$ 0.4) $\times $10$^{-5}$ erg~cm$^{-2}$ (20~keV-10~MeV). The estimated $E_{\rm iso}$ is $3 \times 10^{53}$ erg \citep{2013GCN.15413....1G}. The Nordic Optical Telescope \citep[NOT,][]{2010ASSP...14..211D} equipped with the Alhambra Faint Object Spectrograph and Camera (AlFOSC) instrument provided a redshift of $z = 1.293$.

Due to the significant absorption that photons undergo during propagation, particularly within the host galaxy, optical data not corrected for extinction can only serve as a lower limit on the emission level. In the modelling of the SED of GRB~131030A, optical data were therefore excluded to avoid the considerable uncertainties associated with absorption corrections. Unlike X-rays, where line-of-sight column density provides a relatively robust handle, extinction in the optical band depends strongly on the dust-to-gas ratio and the local environment in the GRB host, both of which are often poorly constrained. Additionally, the presence of a possible thermal component in the optical could further complicate interpretation within the synchrotron+SSC framework, potentially biasing the fit.

 \begin{figure}
     \centering
     \includegraphics[width = 0.45\textwidth]{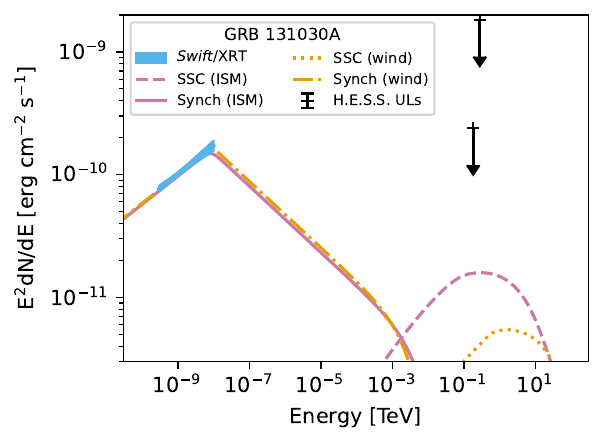}
     \caption{SED of GRB~131030A. The \Swift/XRT spectrum is shown with the blue butterfly. The H.E.S.S. ULs are shown in black. In the plotted emission components, the synchrotron contribution appears as a pink solid curve for the ISM scenario and a yellow dashed-dotted curve for the wind scenario, while the SSC component is represented by a pink dashed curve (ISM) and a yellow dotted curve (wind).}
     \label{fig:modelgrb131030A}
 \end{figure}

The H.E.S.S. observations were grouped into two clusters starting at T$_0$+492\,s for a total of 1440\,s and T$_0$+2244\,s for a total of 2880\,s.  In this section, only the upper limits from the first cluster are considered, as they correspond to earlier times in the GRB afterglow and will provide better constraints on the afterglow emission. For the fit of the \Swift/XRT spectrum, contemporary to the first H.E.S.S. observation cluster, a Galactic column density of $N_{\rm H,gal} = 5.62\times10^{20}\,\mathrm{cm}^{-2}$ was used. The spectral fit provides an intrinsic column density of $N_{\rm H,int}=(1.5\pm0.455)\times 10^{21}\,\mathrm{cm}^{-2}$, and power-law parameters, $\Phi_{0,XRT}= (6.33\pm 0.211)\times10^{-2}$\,keV$^{-1}$\,cm$^{-2}$\,s$^{-1}$ and $\alpha_{\rm XRT} = 1.76 \pm 0.04$.

In Fig.~\ref{fig:modelgrb131030A}, we show the \Swift/XRT spectrum and H.E.S.S. upper limits together with the synchrotron and synchrotron self-Compton modelling. The resulting model parameters are provided in Table~\ref{tab:modelling}. 

\subsubsection{GRB~161001A}

\Swift/BAT initially detected GRB~161001A on October 1, 2016, at 01:05:26 UTC (T$_0$). \Swift/XRT began observing this GRB 65\,s later, while \Swift/UVOT found no significant optical counterpart at similar timescales. The GRB prompt emission phase was also detected by \Fermi/GBM, measuring a T$_{90}$ of 2.2\,s (50-300\,keV). The redshift of this GRB determined by the X-shooter spectrograph, is $z=0.891$~\citep{GCN19971,xshooter_redshift_cat}. The burst has a fluence of $(8 \pm 3.5) \times 10^{-6} \, {\rm erg/cm^{2}}$ detected by the Konus-Wind instrument (20 keV - 10 MeV) \citep{2016GCN.19977....1F}. The corresponding E$_{\rm iso}$ is $2 \times 10^{52}$ erg. 

\begin{figure}
    \centering
    \includegraphics[width = 0.45\textwidth]{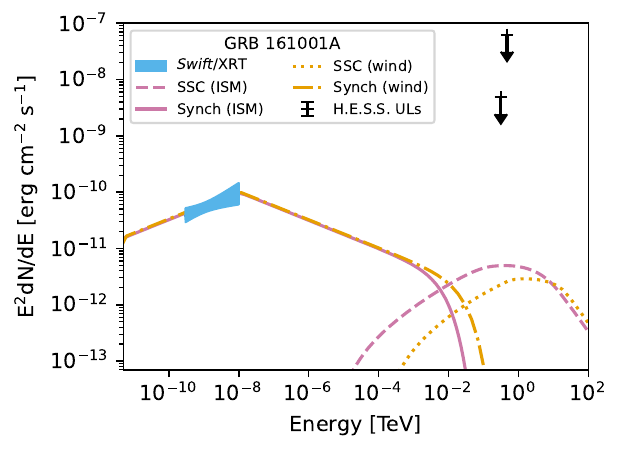}
    \caption{SED of GRB~161001A. The \Swift/XRT spectrum is shown with the blue butterfly. The H.E.S.S. ULs are shown in black. For the modelled emission, synchrotron is depicted by a pink solid line in the ISM case and a yellow dashed-dotted line in the wind case. The SSC component is shown as a pink dashed line for ISM and a yellow dotted line for wind. }
    \label{fig:modelGRB161001A}
\end{figure}

The H.E.S.S. analysis was performed by grouping the observations into two clusters (see Sect.\ref{sec:analysis}). The first cluster of observations began 144\,s after T$_0$ and lasted 432\,s. The second cluster started at 3312\,s after T$_0$ and covered 3960\,s of exposure. As for GRB~131030A, only the first cluster was considered here. 
The spectral fit of the \Swift/XRT data contemporaneous to the H.E.S.S. upper limits of the first cluster results in an intrinsic column density of $N_{\rm H,int}=(1.91\pm0.44)\times 10^{22}$, and power-law parameters $\Phi_{\rm 0,XRT}= (3.29\pm 0.667)\times10^{-2}$\,keV$^{-1}$\,cm$^{-2}$\,s$^{-1} $ and $\alpha_{\rm XRT} = 1.75 \pm 0.17$. For this fit, a Galactic column density of $N_{\rm H,gal} = 1.11\times10^{20}$ was used. The resulting parameters of the single-zone SSC model can be found in Table~\ref{tab:modelling}, and the corresponding spectral energy distribution (SED), including the \textit{Swift}/XRT spectrum, H.E.S.S. upper limits, and modelled emission components, is shown in Fig.~\ref{fig:modelGRB161001A}. Although contemporaneous \Fermi/LAT observations exist, they are not considered, as the energy range covered coincides with the transition between the synchrotron and SSC components and thus no meaningful constraints on the modelling can be provided.

\subsection{Comparing the properties of GRBs observed at VHEs}

This section compares the properties of the GRBs detected at VHE with the GRBs without detections presented in this work and those of the overall population detected by \Swift/BAT, \Swift/XRT and \Fermi/GBM, to look for potential observational biases. GRBs co-detected by \Swift/BAT and \Fermi/GBM (hereby defined as \codetected) were also considered as an additional population for comparing the distributions. 

The criteria for GRB follow-up observations in H.E.S.S. have undergone significant evolution over the years. Additionally, the burst advocate can make observational decisions. Therefore, it is crucial to evaluate potential biases that may be present in our sample. For this, a first comparison between the GRBs observed by H.E.S.S. and the overall GRB sample of the \Fermi and \Swift catalogues was performed to ensure no significant biases in our selections of observed GRBs. GRB 180720B and GRB 190829A were included in the H.E.S.S. sample as they were detected during the period covered in this work.  For the overall sample, the selection was not restricted to the same time interval as this study (i.e. it comprises all GRBs detected by satellite missions throughout the years). After this first check, the distributions of GRBs detected at VHE were compared to the overall GRB catalogue sample used in this study to search for properties specific to VHE-detected GRBs. The distribution comparisons were performed using a Kolmogorov–Smirnov test~\citep[KS test,][]{ks-test}. As GRB~190829A is the only candidate for the low-luminosity GRB class among all the GRBs detected at VHE~\citep{Chand_2020}, the comparison of GRBs detected at VHE was performed twice: including and excluding GRB~190829A.

The KS test checks whether the subset of GRBs observed by H.E.S.S. (and those detected at VHE) is statistically representative of the larger GRB population, thereby identifying any selection biases in our follow-up strategy. To do so, we compared, for each instrument (\Swift/BAT, \Fermi/GBM, and \Swift/XRT) and for redshift, the distributions of key prompt- and afterglow-phase parameters among (1) all catalogued GRBs; (2) those observed by H.E.S.S. (including GRB~180720B and GRB~190829A); and (3) the VHE detections (with and without GRB~190829A). Although every VHE detection occurred during the afterglow, comparing prompt-phase properties (from \Swift/BAT and \Fermi/GBM) could reveal shared characteristics. Prompt-phase values were taken from the \Swift/BAT\footnote{https://swift.gsfc.nasa.gov/results/batgrbcat/}~\citep{2016ApJ...829....7L} and \Fermi/GBM\footnote{https://heasarc.gsfc.nasa.gov/w3browse/fermi/fermigbrst.html}~\citep{2020ApJ...893...46V} catalogues, focusing on \Swift/BAT-detected GRBs due to their precise localisations. In the subsections below, we report KS-test $p$ values for each comparison and highlight any significant deviations.

\subsubsection{Prompt phase: \Swift/BAT}

\begin{figure*}[]
\centering
\includegraphics[width=0.9\textwidth]{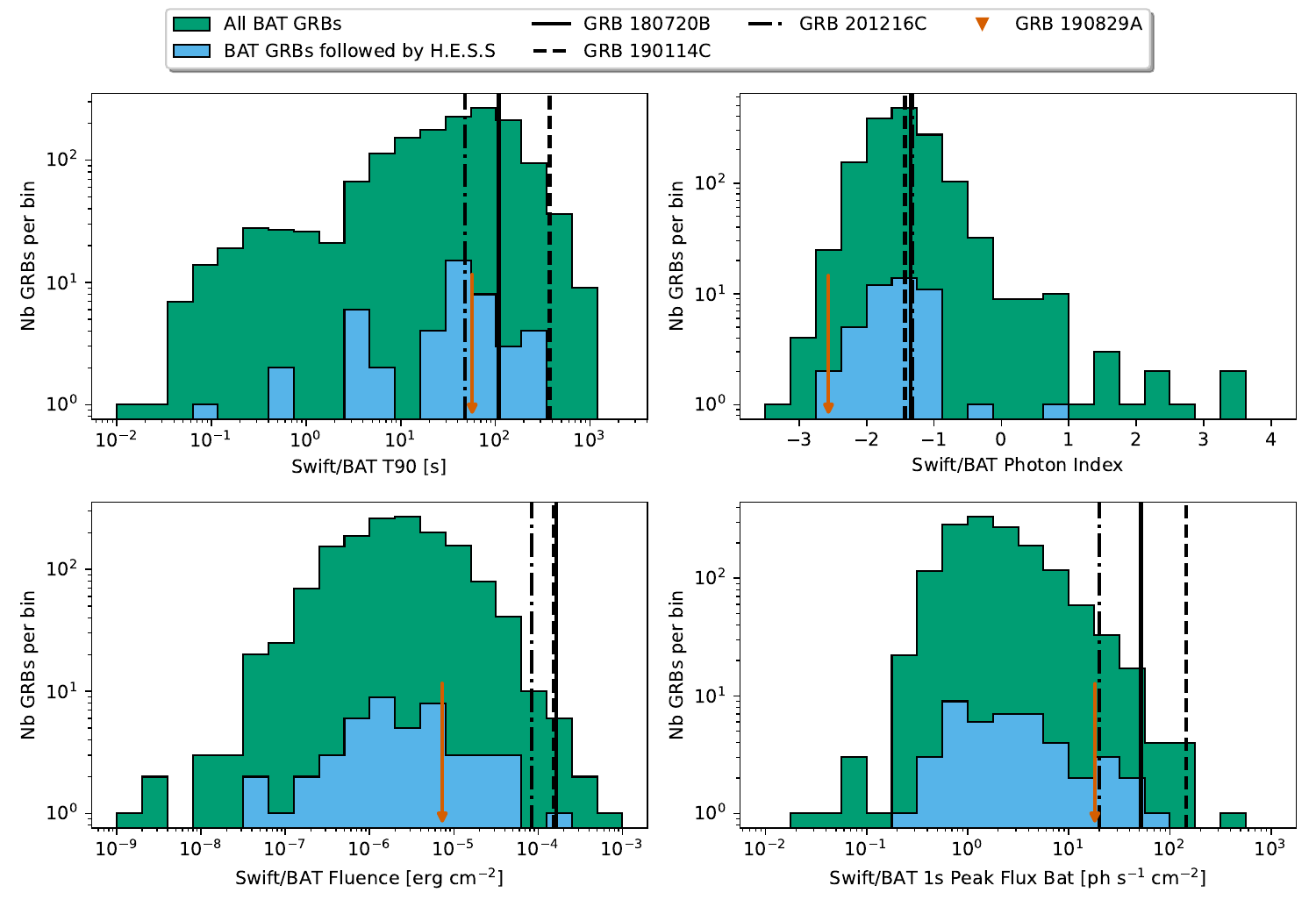}
\caption[]{Distribution of GRB properties provided by \Swift/BAT. The green colour represents the whole \Swift/BAT population, in blue, the one observed by H.E.S.S. The black lines represent the individual measurements for all GRBs detected at VHE and by \Swift/BAT, except for GRB 190829A, represented with an orange arrow. Top-left: Distribution of the $T_{90}$ parameters. Top-right:\ Photon index. Bottom-left:\ Overall fluence of the burst. Bottom-right: One-second peak photon flux.}
\label{fig:population_BAT}
\end{figure*}

Table~\ref{tab:population_BAT} summarises the results of all the KS tests performed on the parameters related to the prompt phase observed by \Swift/BAT. No significant bias was identified between the observed sample by H.E.S.S. and the overall \Swift/BAT sample\footnote{GRB 221009A is not part of this sample because \Swift/BAT did not detect its prompt phase, as the source was outside the instrument's field of view at the time of the event.}. Among the parameters investigated, the one-second peak flux (the highest flux measured in a one-second interval of the burst) with a significance of $1.6\sigma$ is the most substantial difference. Therefore, it was concluded that there were no biases in the selected observation sample. Fig. \ref{fig:population_BAT} shows a comparison between four main parameters measured by \Swift/BAT in the 15-150\,keV energy band: the T$_{90}$ value, the photon index, the fluence, and the one-second peak photon flux.
For most studied parameters, no significant deviation was found between the observed and the parameter distributions of the detected samples. Only the fluence exhibits a considerable deviation, with a $p$ value-equivalent significance of $4.7\sigma$, down to $2.9\sigma$ if GRB 190829A is included. The one-second peak flux shows a strong hint of deviation at $3.9\sigma$, up to $4.5\sigma$ if GRB 190829A is included. 

\begin{table*}
    \centering
        \caption[]{$p$ value from the KS test of the \Swift/BAT parameters.} \label{tab:population_BAT}
    \begin{tabular}{lccc}
    \toprule
    Parameter & $p$ value Observed & $p$ value Detected at VHE & $p$ value Detected at VHE (except GRB 190829A) \\
     & VS Whole population & VS Whole population & VS Whole population  \\
     \midrule
    $T_{90}$ & 0.16 ($1.4\sigma$)& 0.13 ($1.5\sigma$)& 0.25 ($1.2\sigma$)\\
    Fluence & 0.44 ($0.8\sigma$)& \SI{4.3e-3}{} ($2.9\sigma$)& \SI{2.5e-6}{} ($4.7\sigma$)\\
    1s Peak Flux & 0.10 ($1.6\sigma$)& \SI{6.2e-6}{} ($4.5\sigma$)& \SI{1.0e-4}{} ($3.9\sigma$)\\
    Spectral Index & 0.24 ($1.2\sigma$)& 0.58 ($0.6\sigma$)& 0.24 ($1.2\sigma$)\\
    \bottomrule
    \end{tabular}

\tablefoot{The test is performed to compare the distribution of the parameters measured by \Swift/BAT between the overall GRB sample, those GRBs observed by H.E.S.S and those detected at VHE.}
\end{table*}

\subsubsection{Prompt phase: \Fermi/GBM }

\begin{figure*}
\centering
\includegraphics[width=0.9\textwidth]{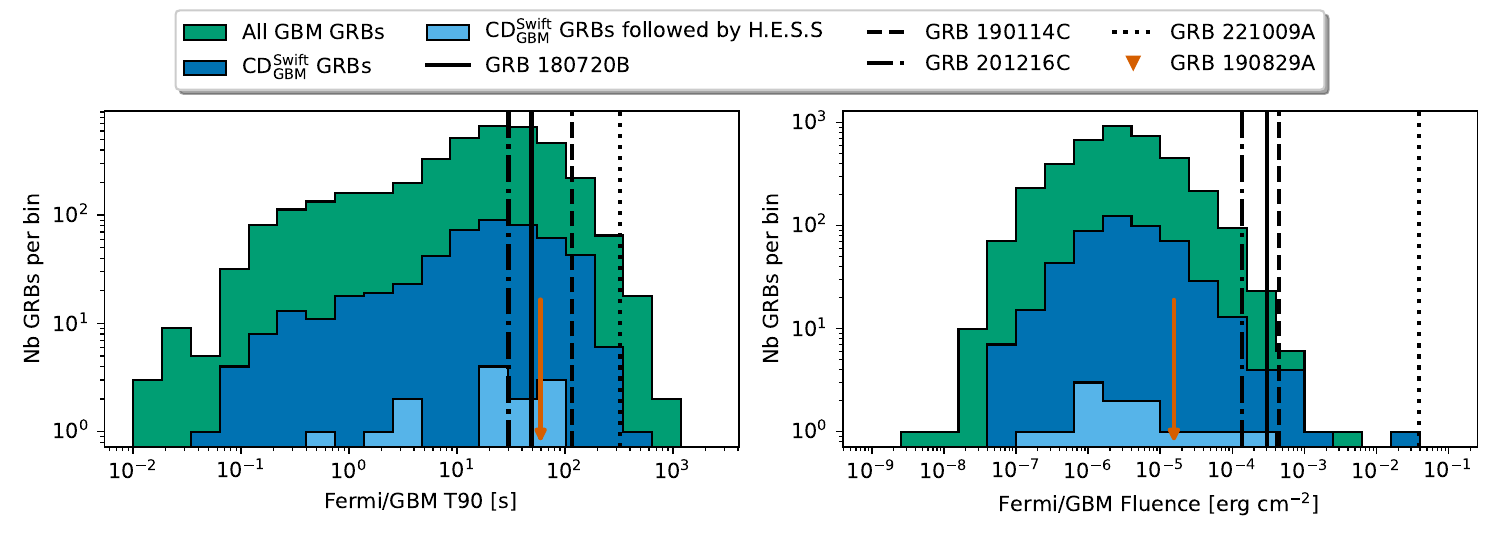}
\caption[]{Distribution of GRB properties measured by \Fermi/GBM. The green colour represents the whole \Fermi/GBM population, in dark blue those detected co-detected by \Fermi/GBM and \Swift/BAT (\codetected) and in light blue the \codetected GRBs observed by H.E.S.S. The black lines represent the individual measurements for all GRBs detected at VHE and by \Fermi/GBM, except for GRB~190829A, which is represented by an orange arrow. Top-left: Distribution of the $T_{90}$ parameters. Top-right plot: Overall fluence of the burst.}
\label{fig:population_GBM}
\end{figure*}

In this section, \Fermi/GBM GRBs observed by H.E.S.S. are considered, but the selection is restricted to those also detected by \Swift/BAT. This allows us to identify more precisely the GRB localisation, therefore ensuring that our observations are covering the true position of the GRB. Similarly to the comparison discussed in the previous section, no significant bias was identified in this set. The correlation study was performed on four parameters measured by \Fermi/GBM (10-1000 keV): the T$_{90}$ value, the fluence, and the peak flux on 64 ms and 1024 ms timescales. No spectral properties were studied, as the values were not available in the online catalogue for most of the GRBs considered. 
Table~\ref{tab:population_GBM} summarises the results of all the KS tests performed for this sample. The results are similar to those obtained for the comparison to the  \Swift/BAT sample, with no significant deviation on T$_{90}$, but with a strong hint of deviation ($> 4\sigma$) on the fluence or the peak flux when excluding GRB 190829A. This is similar to what was observed with \Swift/BAT, therefore indicating that VHE instruments are likely able to detect only very bright bursts. The comparison of the distributions for the T$_{90}$ and fluence is shown in Fig.\ref{fig:population_GBM}.

\begin{table*}
    \centering
        \caption[]{{$p$ value from the KS test comparing the distribution of the measured parameters from   \Fermi/GBM.} \label{tab:population_GBM}
    \begin{tabular}{lccc}
    \toprule
    Parameter & $p$ value Observed & $p$ value Detected at VHE & $p$ value Detected at VHE (except GRB 190829A) \\
     & vs Population \codetected & vs Population \codetected  & vs Population \codetected  \\
     \midrule
    $T_{90}$ & 0.79 ($0.3\sigma$)& \SI{2.7e-2}{} ($2.2\sigma$)& \SI{6.0e-2}{} ($1.9\sigma$)\\
    Fluence & 0.62 ($0.5\sigma$)& \SI{3.8e-4}{} ($3.6\sigma$)& \SI{1.1e-6}{} ($4.9\sigma$)\\
    64 ms Peak Flux & 0.36 ($0.9\sigma$)& \SI{1.6e-5}{} ($4.3\sigma$)& \SI{1.4e-5}{} ($4.3\sigma$)\\
    1024 ms Peak Flux & 0.78 ($0.3\sigma$)& \SI{4.4e-6}{} ($4.6\sigma$)& \SI{8.2e-6}{} ($4.4\sigma$)\\
    \bottomrule
    \end{tabular}

\tablefoot{The test is done between GRBs detected by both \Swift/BAT and \Fermi/GBM, those observed by H.E.S.S., and those detected at VHE. While $T_{90}$ shows no significant deviation, the fluence and both 64~ms and 1024~ms peak fluxes for VHE-detected GRBs differ from the overall population, indicating these bursts tend to be brighter in the prompt phase.}}
\end{table*}

\subsubsection{Afterglow phase: \Swift/XRT}

\begin{figure*}
\centering
\includegraphics[width=0.9\textwidth]{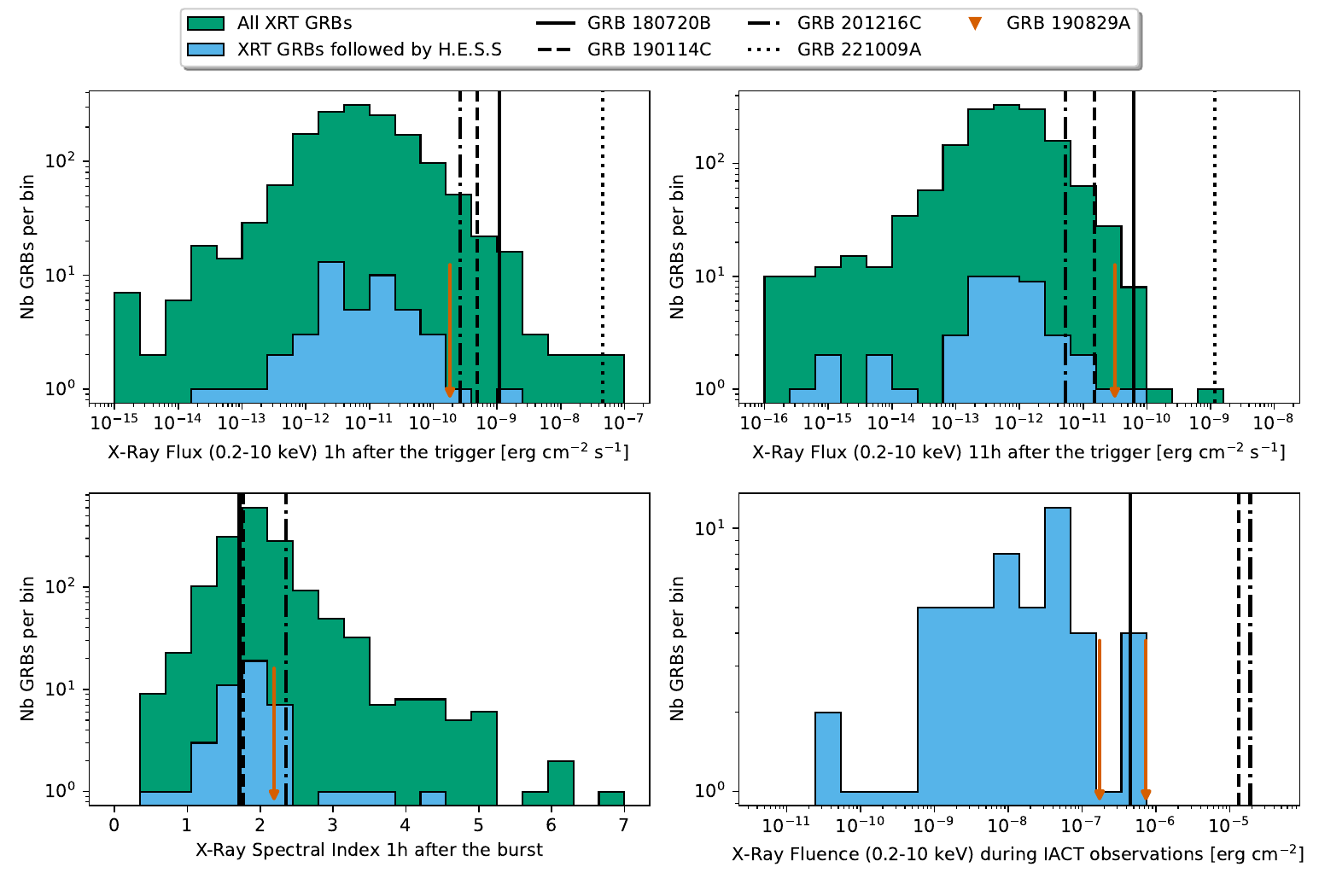}
\caption[]{Distribution of GRB properties measured by \Swift/XRT. Green represents the whole \Swift/XRT population, while blue represents those observed by H.E.S.S. Black lines represent the individual measurements for all GRBs detected at VHE and by \Swift/XRT, except for GRB~190829A, which is represented with the orange arrow. Top-left plot shows the distribution of the $T_{90}$ parameters, and the top-right shows the overall fluence of the burst. The two orange arrows for GRB~190829A in the bottom right panel correspond to the second and first night of H.E.S.S. observations in that order. }
\label{fig:population_XRT}
\end{figure*}

A study of correlations in observables from the afterglow phase with \Swift/XRT is motivated by two key factors. First, all the detections of VHE emission from GRBs have been associated with afterglow emission. Second, some observations reveal similarities between X-ray and VHE emission \citep{HESS_GRB180720B, HESS_GRB190829A}.
The measurements available in the \Swift/XRT live catalogue were used \citep{2007A&A...469..379E, 2009MNRAS.397.1177E, 2023MNRAS.518..174E}, and especially the information on the light curve shape, the flux measured by XRT and the spectral properties measured during each phase of the light curve. The XRT catalogue provides information on the number of breaks in the light curve by fitting power law segments. With this information, the 0.2-10\,keV flux and spectral index at 200\,s, 1\,h, 11\,h and 24\,h were computed (and extrapolated if needed). For the observed and VHE-detected GRBs, the 0.2-10\,keV fluence during the H.E.S.S. observations window was also computed. GRB 221009A was not included for this last parameter, as the VHE detection happened during the rise of the afterglow, while the first measurement performed by \Swift/XRT was performed in a later phase of the afterglow, making it difficult to rely on the extrapolation of the values. The distribution of a few of those parameters can be seen in Fig.~\ref{fig:population_XRT}.
No significant deviations exist in the tested distributions between the GRBs observed and the overall sample. While the comparison between spectral indices does not show any significant deviation between those detected at VHEs and the overall sample, it is possible to see a hint of deviation for the flux and fluence parameters with deviation above 3$\sigma$ for the tested distributions and above 4$\sigma$ for most of them. These results go in a similar direction as those from the prompt phase, the main difference between the ones detected at VHE is their brightness. The results from each KS test are summarised in Table~\ref{tab:population_XRT}.

\begin{table*}
    \caption[]{$p$ value from the KS test of the \Swift/XRT light-curve parameters. } \label{tab:population_XRT}
    \centering
    \begin{tabular}{lccc}
    \toprule
    Parameter & $p$ value observed & $p$ value detected at VHE & $p$ value detected at VHE (except GRB 190829A) \\
     & vs whole population & vs whole population & vs whole population  \\
     \midrule
    Fluence during observations & N/A & \SI{6.5e-5}{} ($4.0\sigma$)& \SI{4.6e-5}{} ($4.1\sigma$)\\
    excluding GRB 221009A &  &  &  \\
    Flux at 200 s & 0.91 ($0.1\sigma$)& \SI{1.9e-3}{} ($3.1\sigma$)& \SI{1.0e-5}{} ($4.4\sigma$)\\
    Flux at 1 h & 0.89 ($0.1\sigma$)& \SI{1.3e-6}{} ($4.0\sigma$)& \SI{6.6e-6}{} ($4.5\sigma$)\\
    Flux at 11 h & 0.92 ($0.1\sigma$)& \SI{9.7e-6}{} ($4.8\sigma$)& \SI{1.1e-4}{} ($3.8\sigma$)\\
    Flux at 24 h & 0.91 ($0.1\sigma$)& \SI{3.7e-5}{} ($4.1\sigma$)& \SI{3.2e-4}{} ($3.6\sigma$)\\
    Spectral index at 200 s & 0.97 ($0.04\sigma$)& 0.60 ($0.5\sigma$)& 0.26 ($1.1\sigma$)\\
    Spectral index at 1 h & 0.48 ($0.7\sigma$)& 0.68 ($0.4\sigma$)& 0.31 ($1.0\sigma$)\\
    Spectral index at 11 h & 0.73 ($0.3\sigma$)& 0.77 ($0.3\sigma$)& 0.82 ($0.2\sigma$)\\
    Spectral index at 24 H & 0.74 ($0.3\sigma$)& 0.75 ($0.3\sigma$)& 0.35 ($0.9\sigma$)\\
    \bottomrule
    \end{tabular}

\tablefoot{The KS test is performed to compare the distribution of the parameters computed from the light curve measured by \Swift/XRT between the overall sample, the ones observed by H.E.S.S and those detected at VHE. Significant deviations ($p<$0.01) are seen for fluence during observations and for flux at various epochs in VHE-detected GRBs, suggesting these brightness parameters differ from the overall population, while spectral indices remain statistically consistent.}
\end{table*}

\subsubsection{Redshift}

\begin{figure}
\centering
\includegraphics[width=0.45\textwidth]{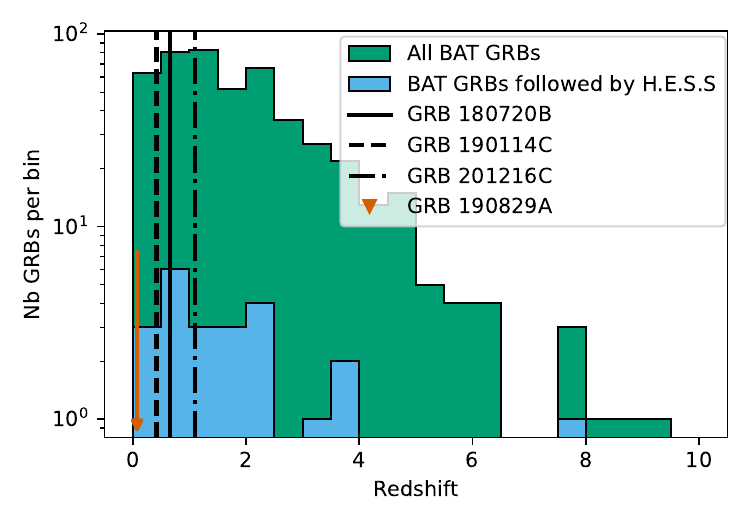}

\caption[]{Distribution of the redshift measured for GRBs detected by \Swift/BAT. The green colour represents the whole \Swift/BAT population with redshift measurements. The redshift distribution of GRBs observed by H.E.S.S. is shown in blue. The black lines represent the individual measurements for all GRBs detected at VHE. The redshift value of GRB~190829A is represented with the orange arrow.}
\label{fig:population_redshift}
\end{figure}

Another critical parameter in detecting GRBs at VHEs is the redshift. Due to the absorption by interaction with the EBL, a significant portion of the gamma-ray emission at VHE is absorbed before reaching the observer. The redshift distribution of the \Swift/BAT sample was compared to that of the GRBs observed by H.E.S.S. and to that of the GRBs detected at VHEs. The distributions are shown in Fig.~\ref{fig:population_redshift}. Both comparisons show a 2$\sigma$ deviation.

\section{Discussion}
\label{sec:discussion}

\begin{figure}
\centering
\includegraphics[width=0.5\textwidth]{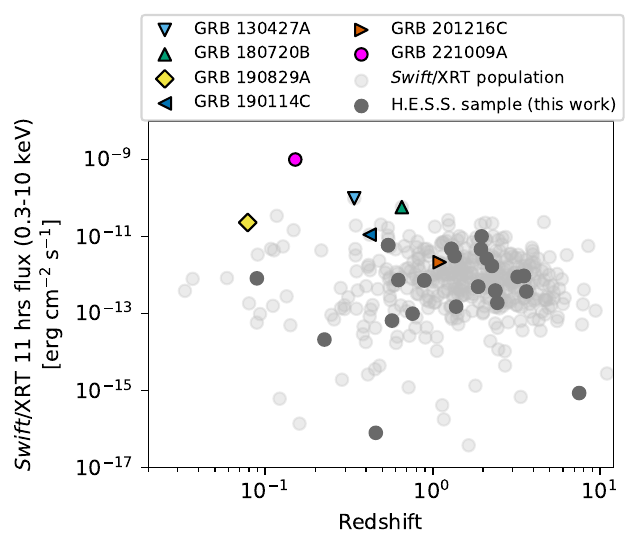}

\caption[]{\Swift/BAT 11-hour flux as a function of the redshift. Light-grey points correspond to all \Swift/GRBs with measured redshift. Dark-grey points correspond to the H.E.S.S. GRBs presented in this work that have a redshift and \Swift/XRT afterglow measurement. The GRBs detected at VHE and GRB~130427A are highlighted in colour. }
\label{fig:XraysRedshift}
\end{figure}

The sample of H.E.S.S. GRBs analysed in this work is broadly consistent with the overall population of \Swift and \Fermi GRBs in terms of redshift and duration. However, the GRBs detected at VHEs exhibit significantly brighter prompt and afterglow emission compared to the general \Swift GRB distribution. This suggests that current IACTs, including H.E.S.S., are primarily sensitive to the most luminous GRBs, and likely miss the majority of events due to limited sensitivity.
 Also, due to the EBL absorption, the detection horizon is limited. To summarise both of those points, Fig.~\ref{fig:XraysRedshift} shows the \Swift/XRT flux at 11 hours after T$_0$ as a function of redshift for the entire population of \Swift/detected GRBs and the H.E.S.S. sample of this work. The VHE-detected GRBs are shown with specific colours and represent the closest and brightest compared to the whole \Swift/XRT population, except for GRB~201216C. GRB~130427A is also included in the figure, although not formally detected at VHE, as it featured an exceptionally bright afterglow and a \textit{Fermi}/LAT photon with energy close to 100\,GeV~\citep{doi:10.1126/science.1242353}.
 The detection of GRB~201216C could still be explained by the smaller zenith angle and small delay of the observations performed by MAGIC combined with the very bright early afterglow (Fig.~\ref{fig:population_XRT}). 
The fact that most of the GRBs observed by H.E.S.S. presented in this work are rather ``ordinary'', combined with the H.E.S.S. observation delays and conditions (e.g. a high zenith angle increases the energy threshold), can explain the non-detection results in this paper. 

The shortest observation delay in this sample is 66\,s and occurred for GRB~180906B. This delay exceeds its prompt emission duration of 18\,s. This is, however, not the fastest follow-up by H.E.S.S. in the years since the time period discussed in this study. The quickest response from the alert system and telescopes was achieved for GRB~191004B at 28\,s, which occurred during moderate moonlight and has not been considered for analysis in this work. The GRB in this sample with the longest T$_{90}$, GRB~160825A (T$_{90}=6.0$\,min), was observed by H.E.S.S. 262\,min after its onset. This highlights the difficulty for IACTs in capturing the prompt phase of GRBs due to the intrinsic delay in satellite alerts \citep[the median delay value for \Fermi/GBM and \Swift alerts is roughly 30\,s and 42\,s respectively,][]{thehessproceedings} that is longer than the typical prompt emission duration, in addition to observational constraints (moonlight presence, telescope slewing time, duty cycle). Serendipitous observations, as in the case of GRB~060602B, are the best chance for IACTs to study the prompt phase of GRBs. The recent detection of GRB~221009A by LHAASO \citep{LHAASOGRB}, whose location was immediately observable at the LHAASO site, demonstrates the importance of the complementarity of Wide Field of View instruments with IACTs. 

Recently, \cite{Ashkar_2024} presented a study that determined potentially detectable GRBs by current IACT observatories by examining observability characteristics and X-ray flux and redshift of archival \Swift and \Fermi/LAT alerts. In the case of H.E.S.S., seven GRBs were identified by the study: GRB~190829A and GRB~180720B, which are not presented here because of dedicated publications, GRB~100621A and GRB~131030A, which are discussed in Sect.~\ref{sec:specificgrbs}, and GRB~060904B, GRB~130925A and GRB~161219B, for which H.E.S.S. did not perform any follow-ups due to bad weather conditions. This indicates that all GRBs identified as potentially detectable by H.E.S.S. were either observed or missed for reasons outside of human control, such as unfavourable weather.

A modelling of three selected GRBs has been presented in this work. This modelling approach does not account for the modification of the KN suppression into the electron-cooling, and therefore does not include the corresponding modifications to the synchrotron spectral shape. However, as seen in the results of these three GRBs, the SSC peak lies well below the synchrotron peak, hence the Klein–Nishina suppression does not significantly affect the electron cooling, and modifications to the synchrotron spectrum are therefore not required. The most constraining upper limits are obtained for GRB~131030A. The SSC scenario discussed in this work can fully explain the non-detections by H.E.S.S. Hard \Swift/XRT spectra from these selected GRBs imply that the emitting electrons are in the slow cooling regime. 
We note that our modelling results just provide one set of feasible parameters for each selected GRBs under the assumption that $\eta_{\rm inj} = 100\%$ and $E_{\rm sh}/E_{\rm iso} =$ 10 or 100. Since degeneracies exist between the parameters, the specific values should not be over-interpreted. For the GRBs modelled here, the SSC component predicted from our afterglow fits lies well below the H.E.S.S. upper limits once Klein–Nishina effects and EBL attenuation are included. The resulting non-detections are therefore expected and indicate no tension with a standard forward-shock scenario. The required sensitivity to challenge our SSC models is still low, suggesting that earlier observations or lower thresholds would be decisive for future studies. In addition to this modelling approach, it was found that a scenario that pushes the SSC flux to reach the H.E.S.S. ULs would require values of E$_{\rm sh}$ at $\mathcal{O} ({10^{55)}}$\,erg, far above cannonical values, and compared only to extreme illustrative cases~\citep{MAGIC_GRB190114C_2}. This consideration reinforces that VHE detections are, in practice, largely unfeasible under present energy thresholds and instrument response times.

Providing more information on the emission above 100 GeV is an important issue in GRB modelling. As shown in the previous section, the lack of multi-wavelength information makes broadband emission modelling difficult for the handful of events whose upper limits at VHE set constraints. The lack of redshift estimation is equally essential:  only approximately 30\% of the GRBs detected in the last decade have a redshift measurement. This is usually due to a lack of follow-ups from optical instruments. 
In November 2022, the H.E.S.S. collaboration set up a public web page to announce the follow-up of GRBs in real time\footnote{\url{https://grbhess.github.io/}} as an effort to motivate deeper and more simultaneous follow-ups at different energies with other instruments.

The lack of detection at VHEs highlights the severe difficulty of overcoming the limitations when observing distant sources due to the EBL absorption. The observations with CT~5 presented here have a lower energy threshold than those carried out with CT~1-4 only (see Table~\ref{tab:analysislocGRBs}). CT~5 also increased the detection prospects of H.E.S.S. by improving the repointing speed of H.E.S.S.

\begin{figure}
\centering
\includegraphics[width=0.5\textwidth]{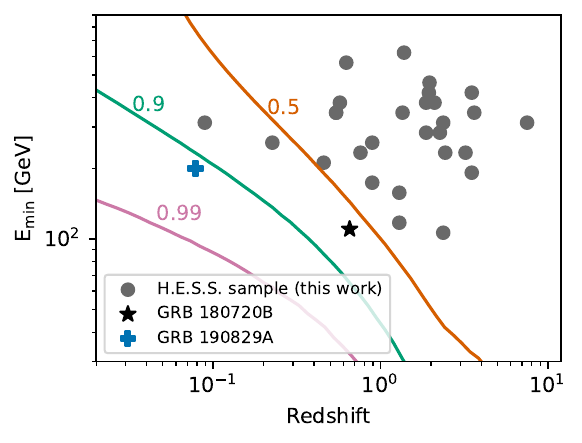}

\caption[]{ Energy threshold and redshift
for the H.E.S.S. sample analysed in this work, shown as grey points. The curves indicate the energy at which, at a given redshift, the EBL absorption coefficient reaches a value of 0.99 (pink curve), 0.9
(turquoise curve), and 0.5 (orange curve). The two VHE-detected GRBs by H.E.S.S. are indicated with a star (GRB~180720B) and blue cross (GRB~190829A).}
\label{fig:ethr_ebl}
\end{figure}

\section{Conclusions and outlook}
\label{sec:conclusions}

This paper presents the H.E.S.S. observations of GRBs over nearly 15 years, including a re-analysis of GRBs observed between 2003 and 2007 using improved calibration and reconstruction algorithms. The GRB selection was performed by searching for coincidences between the H.E.S.S. observations and the publicly available listings of X-ray-detected GRBs. With this method, two additional GRBs could be added to the study, as they occurred within the field of view of another VHE source being observed with H.E.S.S. 

 The data analysis of the GRBs presented in this paper did not yield any detections. Spectral-flux upper limits were provided for GRBs whose localisation uncertainties were smaller than the field of view of H.E.S.S. A stacked analysis was performed on this sample to determine a possible detection in the overall sample, but it did not result in a detection either. Upper-limit maps were obtained for GRBs with larger uncertainties in their localisations. This catalogue of GRB ULs at VHE is the largest released to date.  
A discussion on three specific GRBs was presented with dedicated modelling based on the \textit{Swift}/XRT spectrum contemporaneous with H.E.S.S. observations. These GRBs were selected due to their favourable observation conditions, low redshift, and high X-ray flux. In all three cases, the H.E.S.S. upper limits are consistent with the SSC emission expected from our X-ray afterglow fits and only earlier observations or lower thresholds will allow such models to be critically tested.

The lack of redshift measurements for GRBs poses a challenge to efforts meant to constrain GRB emission models and enable population studies. 
With the recent launch of SVOM\footnote{\url{https://www.svom.eu/en/the-svom-mission/}} and specifically thanks to the satellite’s anti-Solar pointing and dedicated ground-based optical telescopes, it is expected that the availability of redshift estimates will become much more common. 

While current-generation IACTs are limited in sensitivity and energy threshold, future facilities, such as CTAO, are expected to improve these aspects significantly. The lower threshold and faster repositioning capabilities of instruments such as the Large-Sized Telescopes (LSTs) could help address some of the observational challenges highlighted in this study, including delayed follow-up or reduced sensitivity to distant GRBs~\citep{Cortina:2019RL}.

Considering the distribution of \Swift/XRT GRBs in Fig.~\ref{fig:XraysRedshift}, the improved sensitivity (by one order of magnitude) of CTAO compared to H.E.S.S. (excluding differences in the energy threshold) is expected to be able to detect the VHE emission of \Swift/XRT GRBs with an afterglow flux level as low as 10$^{-12}$ erg~cm$^{-2}$~s$^{-1}$ and with redshifts of $z=1$. 
The LSTs will achieve energy thresholds of $\sim$ 30 GeV under optimal observing conditions~\citep{2023ApJ...956...80A}. The EBL absorption shown in Fig.~\ref{fig:ethr_ebl} illustrates the impact of low energy thresholds on the ability to observe distant sources. Considering a scenario where 90\% of the photons are absorbed (i.e. transmission $e^{-\tau} = 0.1$), an energy threshold of 100\,GeV only allows for detections up to $z \sim 0.3$. An energy threshold of 30\,GeV extends the range to $z\sim 0.8$. For distant sources, gamma rays absorbed by the EBL trigger cascades that produce lower-energy secondary photons. Cascades shift energy into the GeV domain but do not replenish the lost TeV photons for high-redshift sources. 

The synergies that will be possible thanks to the presence of the Southern Wide-field Gamma-ray Observatory \citep[SWGO,][]{2023arXiv230904577C} in the southern hemisphere, as well as LHAASO and the High-altitude Water Cherenkov Observatory \citep[HAWC,][]{HAWC:2019xhp} in the northern hemisphere are also highly anticipated. Given their high-duty cycle and survey-mode observation, these wide-field instruments have the capacity to detect GRBs without an external trigger, potentially serving as an alert system for deep observations of a VHE-emitting GRB with IACTs.

\section*{Data availability}
The data for all the tables of this manuscript are available in electronic form at the CDS via anonymous ftp to cdsarc.u-strasbg.fr (130.79.128.5) or via http://cdsweb.u-strasbg.fr/cgi-bin/qcat?J/A+A/.

\begin{acknowledgements}
The support of the Namibian authorities and the University of Namibia in facilitating the construction and operation
of H.E.S.S. is gratefully acknowledged, as is the support by the German Ministry for Education and Research (BMBF), the Max Planck Society, the German Research Foundation (DFG), the Helmholtz Association, the Alexander von Humboldt Foundation, the French Ministry of Higher Education, Research and Innovation, the Centre National de la Recherche Scientifique (CNRS/IN2P3 and CNRS/INSU), the Commissariat à l’énergie atomique et aux énergies alternatives (CEA), the U.K. Science
and Technology Facilities Council (STFC), the Knut and Alice Wallenberg Foundation, the National Science Centre, Poland grant no. 2016/22/M/ST9/00382, the South African Department
of Science and Technology and National Research Foundation, the University of Namibia, the National Commission on Research, Science \& Technology of Namibia (NCRST), the Austrian Federal Ministry of Education, Science and Research
and the Austrian Science Fund (FWF), the Australian Research Council (ARC), the Japan Society for the Promotion of Science
and by the University of Amsterdam. This study was funded by the European Union - NextGenerationEU, in the framework of the PRIN Project "PEACE: Powerful Emission and Acceleration in the most powerful Cosmic Explosion" (code 202298J7KT– CUP G53D23000880006). The views and opinions expressed are solely those of the authors and do not necessarily reflect those of the European Union, nor can the European Union be held responsible for them.  We appreciate the excellent work of the technical support staff in Berlin, Zeuthen, Heidelberg, Palaiseau, Paris, Saclay, Tübingen and in Namibia in
the construction and operation of the equipment. This work benefited from services provided by the H.E.S.S. Virtual Organisation, supported by the national resource providers of the EGI Federation.

All figures were made with the Okabe-Ito colour palette adjusted for colour-blind people.
\end{acknowledgements}

\bibliographystyle{aa}
\bibliography{biblio2}

\clearpage
\makeatletter
\if@longauth
\kern6pt\hrule\kern6pt\institutename
\@longauthfalse
\fi
\makeatother
\clearpage
\newpage
\thispagestyle{empty} 

\begin{appendix}

\section{Additional tables}

\begin{table}[H]
    \centering
\caption[]{GRB population after different selection and analysis stages.} 
    \begin{tabular}{lccc}
        \toprule
        Stage & loc GRBs& un-loc GRBs & Total \\
        \midrule
        
        Follow-up observations & - & - & 107 \\
        Selected for analysis & 66 & 23 & 89 \\
        Retained after quality selection & 48 & 15 & 63 \\
        Flux ULs determined & 48 & 1 & 49 \\ 
        
        \bottomrule
    \end{tabular}

\label{tab:grb_population_analysis}

\tablefoot{GRBs are classified as well-localised (loc, typically \textit{Swift}/BAT alerts with position uncertainty $\lesssim$3$'$) or poorly localised (un-loc, mostly \textit{Fermi}/GBM alerts with uncertainty $\gtrsim$0.2$^\circ$).}
\end{table}

\begin{table}[H]
    \centering
        
    \caption[]{H.E.S.S. Follow-up observations triggered by \Fermi/GBM.  \label{tab:analysisGRBsGBM}}
    \resizebox{0.5\textwidth}{!}{\centering
    \begin{tabular}{lccc} 
         \toprule
          Name & $T_{\rm 0}$ & delay (min)      & Analysis configuration \\
          \midrule
GRB~150127589  &  2015-01-27 14:08:26  &  588.5  &  stereo loose \\
GRB~150422703  &  2015-04-22 16:52:33  &  188.7  &  hybrid loose \\
GRB~160113398  &  2016-01-13 09:32:30  &  997.2  &  stereo loose \\
GRB~160822672  &  2016-08-22 16:07:40  &  144.1  &  hybrid loose \\
GRB~160825799  &  2016-08-25 19:10:49  &  94.5  &  mono loose \\
GRB~160825799  &    &  205.4  &  mono loose \\
GRB~170402961  &  2017-04-02 23:03:25  &  262.5  &  stereo loose \\
GRB~170730133  &  2017-07-30 03:11:44  &  1206.9  &  hybrid loose \\
GRB~170826819  &  2017-08-26 19:38:56  &  92.4  &  hybrid loose \\
GRB~171112868  &  2017-11-12 20:50:17  &  22.4  &  hybrid loose \\
GRB~180522607  &  2018-05-22 14:34:38  &  737.4  &  hybrid loose \\
GRB~180906759  &  2018-09-06 18:12:25  &  1.1  &  hybrid loose \\
GRB~190306943  &  2019-03-06 22:37:43  &  84.5  &  stereo loose \\
GRB~190306943  &     &  237.3  &  stereo loose \\
GRB~190507970  &  2019-05-07 23:16:29  &  29.0  &  hybrid loose \\
GRB~190727668  &  2019-07-27 16:01:52  &  107.4  &  hybrid loose \\
\bottomrule
    \end{tabular}
}
\tablefoot{The \Fermi/GBM name identifier is used here for uniformity as many of these GRBs do not have an official GCN name. The second column corresponds to the time of the onset of the burst (T$_0$), the third column indicates the delay of the H.E.S.S. observation in minutes and the fourth column indicates the analysis configuration (see Sect.~\ref{sec:analysis}). Columns without T$_0$ values correspond to subsequent H.E.S.S. observations of the GRB in the previous column with provided T$_0$. In Sect.~\ref{sec:analysis}, the results of UL maps for these GRBs are detailed. }
     \end{table}
     
\begin{table*}
\centering
    \caption[]{Multi-wavelength information from the GCN Circulars for our GRB sample (loc and un-loc). \label{tab:observationGRBs}}
\begin{threeparttable}

\resizebox{0.67\textwidth}{!}{\centering
 \begin{tabular}{llccccccccl}
\toprule
Name & Instrument & RA (J2000)$^a$ & Dec (J2000)$^a$ & error & T$_{90}^b$ & R$^c$ & O$^c$ & HE$^c$ & z \\
     &            &  &  & (arcmin) & (s)   &    &   &    &   \\
     \midrule 
    
GRB~041006 & HETE-2 & 00h54m53s & +01d12m04s & 5.00 &  $\sim$20.0 & $\checkmark$ & $\checkmark$ & $\times$ & 0.716\tnote{1}\\
GRB~041211B$^p$ & HETE-2 & 06h43m12s & +20d23m42s & 1.33 &  $>$100.0 & $\cdot$ & $\times$ & $\times$ & - \\
GRB~050209 & HETE-2 & 08h26m09s & +19d41m02s & 14.00 &  46.0 & $\cdot$ & $\times$ & $\cdot$ & - \\
GRB~050509C & HETE-2 &  12h52m54s &  -44d50m04s & 0.02 &  25.0 & $\checkmark$ & $\checkmark$ & $\cdot$ & - \\
GRB~050607 & \Swift/BAT &  20h00m43s &  +09d08m20s & 0.13 & 26.4 & $\cdot$ & $\checkmark$ & $\cdot$ & - \\
GRB~050726 & \Swift/BAT &  13h20m06s &  -32d03m50s & 0.10 & 49.9 & $\cdot$ & $\checkmark$ & $\cdot$ & - \\
GRB~050801 & \Swift/BAT &  13h36m35s & -21d55m48s & 0.02 & 19.4 & $\times$ & $\checkmark$ & $\cdot$ & 1.56\tnote{2}\\
GRB~060505 & \Swift/BAT &  22h07m05s & -27d48m54s & 0.08 & 4.0 & $\cdot$ & $\checkmark$ & $\cdot$ & 0.089\tnote{3} \\
GRB~060526 & \Swift/BAT &  15h31m20s &  +00d17m17s & 0.11 & 298.2 & $\cdot$ & $\checkmark$ & $\cdot$ & 3.21\tnote{4} \\
GRB~060728 & \Swift/BAT &  01h06m35s & -41d23m24s & 3.00 & 60.0 & $\cdot$ & $\times$ & $\cdot$ & - \\
GRB~061110A & \Swift/BAT &  22h25m08s &  -02d15m07s & 0.06 & 40.7 & $\cdot$ & $\checkmark$ & $\cdot$ & 0.758\tnote{5} \\
GRB~070209 & \Swift/BAT &  03h04m51s &  -47d22m34s & 2.80 &  0.1 & $\cdot$ & $\times$ & $\cdot$ & 0.314\tnote{6} \\
GRB~070419B & \Swift/BAT &  21h02m50s &  -31d15m58s & 3.94 & 236.4 & $\cdot$ & $\checkmark$ & $\cdot$ & - \\
GRB~070612B & \Swift/BAT &  17h26m52s &  -08d44m49s & 0.08 & 13.5 & $\cdot$ & $\times$ & $\cdot$ & - \\
GRB~070621 & \Swift/BAT &  21h35m13s &  -24d48m32s & 0.03 & 33.3 & $\cdot$ & $\times$ & $\cdot$ & - \\
GRB~070721A & \Swift/BAT &  00h12m35s & -28d31m48s & 0.04 & 3.4 & $\cdot$ & $\checkmark$ & $\cdot$ & - \\
GRB~070721B & \Swift/BAT &  02h12m31s &  -02d11m53s & 0.01 & 340.0 & $\times$ & $\checkmark$ & $\cdot$ & 3.63\tnote{7} \\
GRB~070724A & \Swift/BAT &  01h51m18s & -18d36m36s & 0.04 & 0.4 & $\times$ & $\times$ & $\cdot$ & 0.457\tnote{8} \\
GRB~070808 & \Swift/BAT &  00h27m03s & +01d10m48s & 0.03 & 32.0 & $\cdot$ & $\checkmark$ & $\cdot$ & - \\
GRB~070920B & \Swift/BAT &  00h00m30s &  -34d50m38s & 0.13 & 20.2 & $\cdot$ & $\times$ & $\cdot$ & - \\
GRB~071003 & \Swift/BAT\tnote{a}&  20h07m26s &  +10d57m14s & 0.10 & 150.0 & $\checkmark$ & $\checkmark$ & $\cdot$ & 1.604\tnote{9} \\
GRB~080413A & \Swift/BAT &  19h09m12s &  -27d40m37s & 0.01 & 46.0 & $\times$ & $\checkmark$ & $\cdot$ & 2.433\tnote{10} \\
GRB~080804 & \Swift/BAT & 21h54m42s &  -53d11m20s & 3.00 & 34.0 & $\times$ & $\checkmark$ & $\cdot$ & 2.2045\tnote{11} \\
GRB~081221 & \Swift/BAT &  01h03m12s &  -24d32m31s & 0.02 & 34.0 & $\checkmark$ & $\checkmark$ & $\cdot$ & - \\
GRB~081230 & \Swift/BAT &  02h29m19s & -25d08m42s & 0.03 & 60.7 & $\cdot$ & $\checkmark$ & $\cdot$ & 1.28 \tnote{12}\\
GRB~090201 & \Swift/BAT &  06h08m12s &  -46d36m14s & 0.06 & 83.0 & $\times$ & $\checkmark$ & $\cdot$ & 2.1 \tnote{11}\\
GRB~091018 & \Swift/BAT &  02h08m46s &  -57d32m46s & 0.06 & 4.4 & $\cdot$ & $\checkmark$ & $\cdot$ & 0.971 \tnote{11}\\
GRB~100418A & \Swift/BAT &  17h05m26s &  +11d27m25s & 0.03 & 7.0 & $\checkmark$ & $\checkmark$ & $\cdot$ & 0.6239 \tnote{11}\\
GRB~100621A & \Swift/BAT &  21h01m14s &  -51d06m07s & 0.03 & 63.6 & $\cdot$ & $\checkmark$ & $\cdot$ & 0.542 \tnote{11}\\
GRB~110625A & \Swift/BAT &  19h07m00s & +06d45m18s & 0.04 & 44.5 & $\cdot$ & $\checkmark$ & $\checkmark$ & - \\
GRB~120328A & \Swift/BAT &  16h06m26s &  -39d19m19s & 0.03 & 24.2 & $\cdot$ & $\times$ & $\cdot$ & - \\
GRB~130502A & \Swift/BAT &  09h14m19s &  -00d08m02s & 0.03 & 3.0 & $\cdot$ & $\checkmark$ & $\times$ & - \\
GRB~131030A & \Swift/BAT &  23h00m18s & -05d22m48s & 0.08 & 41.1 & $\cdot$ & $\checkmark$ & $\cdot$ & 1.293 \tnote{13}\\
GRB~131202A & \Swift/BAT &  22h56m01s & -21d39m00s & 2.60 & 30.4 & $\cdot$ & $\checkmark$ & $\cdot$ & - \\
GRB~140818B & \Swift/BAT &  18h04m40s &  -01d21m14s & 2.10 & 18.1 & $\cdot$ & $\checkmark$ & $\cdot$ & - \\
GRB~141004A & \Swift/BAT\tnote{a} &  05h06m53s &  +12d49m41s & 1.00 &  3.9 & $\cdot$ & $\checkmark$ & $\cdot$ & 0.571\tnote{14}\\
GRB~150127B$^{\#}$ & \Fermi/GBM &  09h29m38s &  -03d08m24s & 60.00 &  60.9 & $\cdot$ & $\cdot$ & $\cdot$ & - \\
GRB~150301A & \Swift/BAT &  16h17m07s &  -48d43m55s & 1.90 &  0.5 & $\cdot$ & $\cdot$ & $\cdot$ & - \\
GRB~150422A$^{\#}$ & \Fermi/GBM &  14h20m24s & -20d51m36s & 64.20 &  36.9 & $\cdot$ & $\cdot$ & $\cdot$ & - \\
GRB~150711A & \Swift/BAT &  14h46m30s &  -35d27m50s & 1.00 & 64.2 & $\cdot$ & $\times$ & $\cdot$ & - \\
GRB~160113A$^{\#}$ & \Fermi/GBM &  12h29m02s &  +11d31m48s & 72.00 &  24.6 & $\cdot$ & $\cdot$ & $\cdot$ & - \\
GRB~160310A & Fermi/LAT &  06h35m17s &  -07d12m54s & 6.00 & 18.2 & $\cdot$ & $\checkmark$ & $\checkmark$ & - \\
GRB~160712A & \Swift/BAT &  20h16m40s & -26d57m54s & 1.30 &  $\sim$25.0 & $\cdot$ & $\times$ & $\cdot$ & - \\
GRB~160822A$^{\#}$ & \Fermi/GBM &  18h08m31s &  +03d35m02s & 32.40 &  0.0 & $\cdot$ & $\cdot$ & $\cdot$ & - \\
GRB~160825A*$^{\#}$ & \Fermi/GBM &  21h58m10s &  +08d09m36s & 360.00 &  6.1 & $\cdot$ & $\cdot$ & $\cdot$ & - \\
GRB~161001A & \Swift/BAT &  04h47m42s &  -57d15m40s & 1.00 & 2.6 & $\cdot$ & $\checkmark$ & $\cdot$ & 0.891\tnote{15} \\
GRB~170402B*$^{\#}$ & \Fermi/GBM &  20h31m41s & -45d55m48s & 333.00 &  22.5 & $\cdot$ & $\cdot$ & $\cdot$ & - \\
GRB~170531B & \Swift/BAT &  19h07m36s &  -16d24m50s & 1.50 &  164.1 & $\cdot$ & $\checkmark$ & $\cdot$ & 2.366\tnote{16} \\
GRB~170730B*$^{\#}$ & \Fermi/GBM &  21h35m38s & -29d45m00s & 253.80 &  6.7 & $\cdot$ & $\cdot$ & $\cdot$ & - \\
GRB~170826B$^{\#}$ & \Fermi/GBM &  21h50m48s & -31d48m00s & 60.00 &  11.0 & $\cdot$ & $\cdot$ & $\cdot$ & - \\
GRB~171020A & \Swift/BAT &  02h37m02s &  +15d11m56s & 1.80 & 41.9 & $\cdot$ & $\checkmark$ & $\cdot$ & 1.87\tnote{17} \\
GRB~171112A$^{\#}$ & \Fermi/GBM &  01h21m52s & -59d40m41s & 10.20 &  302.8 & $\cdot$ & $\cdot$ & $\cdot$ & - \\
GRB~180510A & \Swift/BAT &  18h25m20s &  -31d55m01s & 1.00 & 40.4 & $\cdot$ & $\checkmark$ & $\cdot$ & - \\
GRB~180512A & \Swift/BAT &  13h27m46s &  +21d24m14s & 1.90 & 24.0 & $\cdot$ & $\cdot$ & $\cdot$ & - \\
GRB~180522A*$^{\#}$ & \Fermi/GBM & 20h00m24s & -16d31m48s &  306.00 & 6.9 & $\cdot$ & $\cdot$ & $\cdot$ & - \\
GRB~180613A & \Swift/BAT &  14h06m11s &  -43d04m55s & 1.80 & 50.8 & $\cdot$ & $\checkmark$ & $\cdot$ & - \\
GRB~180906B*$^{\#}$ & \Fermi/GBM & 18h00m28s & -67d40m12s &  197.40 & 13.3 & $ \cdot$ & $\cdot$ & $\cdot$ & - \\
GRB~190306A*$^{\#}$ & \Fermi/GBM &  15h24m17s &  -00d22m48s & 153.00 &  180.5 & $\cdot$ & $\cdot$ & $\cdot$ & - \\
GRB~190507B$^{\#}$ & \Fermi/GBM &  19h11m17s & -22d49m12s & 71.40 &  36.4 & $\cdot$ & $\times$ & $\cdot$ & - \\
GRB~190627A & \Swift/BAT &  16h19m22s &  -05d18m07s & 2.70 & 1.6 & $\cdot$ & $\checkmark$ & $\cdot$ & 1.942\tnote{18} \\
GRB~190727A$^{\#}$ & \Fermi/GBM &  14h57m58s &  +19d26m24s & 60.00 &  47.4 & $\cdot$ & $\cdot$ & $\cdot$ & - \\
GRB~190821A & \Swift/BAT &  16h40m17s &  -34d01m37s & 1.30 & 57.1 & $\cdot$ & $\checkmark$ & $\cdot$ & - \\
GRB~191004B & \Swift/BAT & 03h16m48s &  -39d38m13s & 1.00 & 37.7 & $\cdot$ & $\checkmark$ & $\cdot$ & 3.503\tnote{19} \\
\bottomrule
    \end{tabular}
    }
\tablefoot{The GRB coordinates reported by the corresponding instrument are listed with the localisation uncertainty, the burst duration (T$_{90}$, measured as the time in which 90\% of the emission is detected), the detection at other energies, and the redshift.
$^a$ For GRBs detected by \textit{Swift}-BAT, the coordinates are obtained from \url{https://swift.gsfc.nasa.gov/archive/grb_table/}. For those detected by \textit{Fermi}-GBM, the coordinates are taken from \cite{von_Kienlin_2020}.
For GRBs detected by HETE-2, the coordinates are retrieved from the GCN Circulars.
$^b$ For sources detected by HETE-2, T$_{90}$ values are taken from \cite{HESS_GRB_cat_2009}. 
$^c$ R, O, and HE represent radio, optical, and high-energy (100\,MeV to 100\,GeV) gamma-ray observations. A $\checkmark$ indicates a counterpart detection, $\times$ indicates no detection, and $\cdot$  means no information is available via GCN.
$^p$ This burst is referred to with the suffix ``B'', contrary to the name given by the GCN Circulars, to distinguish it from GRB~041211A, which occurred earlier on the same day (P\'elangeon, private communication).
$^{\#}$ GRBs detected by \Fermi/GBM are not included in Table~\ref{tab:analysislocGRBs}. GRBs with an * after the name indicate that the GRB does not appear with a GCN-style name in the literature, adopting the auto-generated name of GCNweb:~\url{https://user-web.icecube.wisc.edu/~grbweb_public/index.html}. 
}

\tablebib{
(1) \cite{Stanek_2005} 
(2) \cite{2007MNRAS.377.1638D} 
(3) GCN: \cite{GCN5123} 
(4) \cite{GRB060526-redshift} 
(5) GCN: \cite{2007GCN..6759....1F}
(6) GCN: \cite{2007GCN..6101....1B} 
(7) \cite{2011ApJ...731..103X} 
(8) GCN: \cite{2007GCN..6665....1C} 
(9) \cite{2008ApJ...688..470P} 
(10) GCN: \cite{2008GCN..7602....1T} 
(11) \cite{2015A&A...581A.125K} 
(12) \cite{2011A&A...526A.153K} 
(13) GCN: \cite{2013GCN.15407....1X} 
(14) GCN: \cite{2014GCN.16902....1D} 
(15) \cite{xshooter_redshift_cat} 
(16) GCN: \cite{2017GCN.21177....1D} 
(17) GCN: \cite{2017GCN.22039....1M} 
(18) GCN: \cite{2019GCN.24916....1J} 
(19) GCN: \cite{2019GCN.25956....1D} 
}

\end{threeparttable}
\end{table*}

\begin{table*}
     \caption[]{GRB analysis results.} \label{tab:analysislocGRBs}
    \centering
    \resizebox{0.9\textwidth}{!}{
    \begin{tabular}{lccccccccccc} 
         \toprule
          Name & $T_{\rm 0}$ & delay & Exposure      & Analysis configuration & $N_{ON}$ & $N_{OFF}$ & $\alpha_\mathrm{exp}$  & Significance & $E_{th}$  & Flux ULs & Flux ULs \\
                   &  &  &&   &  &  &   &  &  & $\alpha=-2.5$ & $\alpha=-5.0$  \\

        &      &(min)    &(h)     &     &     &      &    &  ($\sigma$) & (GeV) & ($10^{-12}{\rm cm}^{-2}{\rm s}^{-1}$) & ($10^{-12}{\rm cm}^{-2}{\rm s}^{-1}$)\\
          \midrule
GRB~041006  & 2004-10-06 12:18:08  &  626.6 &  1.2  &  stereo loose  &  48  &  467  &  8.79  &  -0.67  &  232  &  2.3 & 3.6 \\
GRB~041211B  &  2004-12-11 11:31:46  &  567.8 &  1.7  &  stereo loose  &  65  &  806  &  14.73  &  1.3  &  463  &  2.8 & 4.1\\
GRB~050209  &  2005-02-09 01:31:41  &  1209.3  &  2.5  &  stereo loose  &  91  &  778  &  9.06  &  0.52  &  511  &  2.1 & 2.9\\
GRB~050509C  & 2005-05-09 22:45:54  &  1257.4&  0.8  &  stereo loose  &  30  &  265  &  8.63  &  -0.11  &  257  &  3.6 & 5.2\\
GRB~050607  &   2005-06-07 09:11:23  &  1023.6  &  0.8  &  stereo loose  &  29  &  453  &  14.22  &  -0.49  &  283  &  2.0 & 3.1\\
GRB~050726  &  2005-07-26 05:00:18  &  773.6  &  1.7  &  stereo loose  &  56  &  488  &  9.02  &  0.24  &  283  &  6.3 & 11.2 \\
GRB~050801  & 2005-08-01 18:28:02  &  16.1  &  0.4  &  stereo loose  &  10  &  112  &  8.99  &  -0.68  &  622  &  6.6 & 3.0\\
GRB~060505  & 2006-05-05 06:36:01  &  1163.9 &  1.6  &  stereo loose  &  33  &  368  &  11.23  &  0.04  &  312  &  6.5 & 10.1\\
GRB~060526  &   2006-05-26 16:28:30  &  285.1  &  1.7  &  stereo loose  &  97  &  805  &  8.9  &  0.64  &  232  &  1.7 & 2.4 \\
GRB~060728  &  2006-07-28 22:24:31  &  249.9 &  0.4  &  stereo loose  &  11  &  148  &  9.99  &  -0.99  &  260  &  2.1 & 3.4\\
GRB~061110A  & 2006-11-10 11:47:21  &  408.6 &  1.6  &  stereo loose  &  56  &  447  &  8.11  &  0.11  &  232  &  3.4 & 4.6 \\
GRB~070209  &  2007-02-09 03:33:42  &  926.8  &  0.8  &  stereo loose  &  29  &  234  &  9.1  &  0.6  &  380  &  4.7 & 8.0\\
GRB~070419B  & 2007-04-19 10:44:06  &  907.2  &  0.8  &  stereo loose  &  21  &  154  &  9.18  &  0.94  &  463  &  6.3 & 5.3\\
GRB~070612B  & 2007-06-12 06:21:18  &  900.9 &  1.7  &  stereo loose  &  91  &  802  &  9.07  &  0.26  &  211  &  3.8 & 6.1\\
GRB~070621  &  2007-06-21 23:17:39  &  6.7  &  3.6  &  stereo loose  &  169  &  1410  &  8.71  &  0.53  &  211  &  2.9 & 4.5 \\
GRB~070621  &    &    37.2  &  0.4  &  stereo loose  &  6  &  109  &  9.08  &  -1.83  &  463  &  6.7 & 2.5\\
GRB~070721A  & 2007-07-21 10:01:08  &  833.4  &  1.7  &  stereo loose  &  56  &  822  &  14.22  &  -0.22  &  257  &  5.6 & 8.5 \\
GRB~070721B  &  2007-07-21 10:33:48  &  925.8    &  1.3  &  stereo loose  &  40  &  522  &  13.85  &  0.36  &  345  &  4.8 & 6.1\\
GRB~070724A  &  2007-07-24 10:53:50  &  893.6 &  1.7  &  stereo loose  &  72  &  617  &  8.99  &  0.38  &  211  &  5.7 & 9.0 \\
GRB~070808  &  2007-08-08 18:28:00  &  306.3  &  1.7  &  stereo loose  &  51  &  501  &  9.04  &  -0.57  &  283  &  1.9  & 1.9\\
GRB~070920B  &  2007-09-20 21:04:32  &  221.1 &  1.3  &  stereo loose  &  14  &  248  &  9.28  &  -2.59  &  380  &  2.7 & 1.1 \\
GRB~071003  &  2007-10-03 07:40:55  &  631.5 &  1.5  &  stereo loose  &  39  &  376  &  8.37  &  -0.85  &  345  &  1.9 & 2.5 \\
GRB~080413A  & 2008-04-13 02:54:19  &  35.3  &  0.3  &  stereo loose  &  17  &  104  &  9.05  &  1.43  &  232  &  11.2 & 1.6 \\
GRB~080804  &   2008-08-04 23:20:14  &  6.0  &  0.4  &  stereo loose  &  12  &  102  &  9.06  &  0.21  &  312  &  6.3  & 9.9\\
GRB~080804  &   &  36.2   &  1.3  &  stereo loose  &  31  &  287  &  8.94  &  -0.18  &  312  &  2.0 & 3.5\\
GRB~081221  &  2008-12-21 16:21:11  &  167.7  &  0.8  &  stereo loose  &  34  &  309  &  9.25  &  0.1  &  283  &  2.8 & 3.1\\
GRB~081230  &  2008-12-30 20:36:12  &  70.0  &  0.8  &  stereo loose  &  17  &  214  &  9.09  &  -1.35  &  686  &  3.2 & 1.4\\
GRB~090201  &  2009-02-01 17:47:02  &  214.3  &  0.8  &  stereo loose  &  15  &  185  &  8.93  &  -1.25  &  380  &  2.4 & 2.6\\
GRB~091018  &  2009-10-18 20:48:19  &  77.4  &  1.7  &  stereo loose  &  23  &  253  &  9.09  &  -0.89  &  380  &  1.1& 1.7\\
GRB~100418A  & 2010-04-18 21:10:08  &  133.0 &  0.8  &  stereo loose  &  13  &  134  &  8.87  &  -0.52  &  563  &  4.9 & 2.9 \\
GRB~100621A  &   2010-06-21 03:03:32  &  11.5 &  0.4  &  stereo loose  &  12  &  104  &  8.73  &  0.02  &  345  &  5.2 & 5.7 \\
GRB~100621A  &    &  42.0  &  0.4  &  stereo loose  &  21  &  125  &  9.97  &  2.05  &  345  &  5.7 & 9.9\\
GRB~110625A  &  2011-06-25 21:08:28  &  21.2 &  1.1  &  stereo loose  &  63  &  279  &  5.46  &  1.47  &  283  &  5.9 & 11.7 \\
GRB~120328A  &  2012-03-28 03:06:19  &  6.2  &  0.4  &  stereo loose  &  13  &  145  &  9.03  &  -0.75  &  232  &  3.6 & 3.3 \\
GRB~130502A  &  2013-05-02 17:50:30  &  64.9 &  1.7  &  stereo loose  &  47  &  421  &  9.22  &  0.19  &  312  &  1.8 & 1.8\\
GRB~131030A  & 2013-10-30 20:56:18  &  8.2  &  0.4  &  mono loose  &  66  &  373  &  7.4  &  1.96  &  117  &  11.0 & 17.1 \\
GRB~131030A  &   & 37.4  &  0.8  &  mono loose  &  56  &  403  &  8.32  &  1.0  &  157  &  48.8 & 86.3 \\
GRB~131202A  &  2013-12-02 15:12:09  &  234.9  &  0.8  &  stereo loose  &  23  &  189  &  8.89  &  0.35  &  312  &  6.0& 6.7\\
GRB~140818B  &  2014-08-18 18:44:16  &  1.9  &  0.4  &  mono loose  &  27  &  238  &  8.38  &  -0.24  &  117  &  29.5 & 27.6 \\
GRB~140818B  &    &  60.3   &  0.9  &  mono loose  &  46  &  340  &  8.51  &  0.88  &  161  &  17.1 & 20.1 \\
GRB~141004A  & 2014-10-04 23:20:54  &  160.1 &  1.0  &  stereo loose  &  15  &  211  &  9.41  &  -1.59  &  380  &  1.1 & 1.7 \\
GRB~150301A  &  2015-03-01 01:04:28  &  25.5  &  1.7  &  mono loose  &  73  &  480  &  8.35  &  1.85  &  142  &  20.7 & 37.9 \\
GRB~150711A  &  2015-07-11 18:23:03  &  23.8  &  1.3  &  mono loose  &  44  &  692  &  14.33  &  -0.6  &  177  &  6.7 & 6.3 \\
GRB~160310A  &  2016-03-10 00:22:58  &  1088,1  &  1.3  &  stereo loose  &  44  &  360  &  9.23  &  0.74  &  257  &  4.3 & 5.2  \\
GRB~160712A  &  2016-07-12 19:53:36  &  229.6 &  1.9  &  mono loose  &  235  &  2120  &  8.32  &  -1.17  &  161  &  9.2 & 10.1\\
GRB~161001A  &  2016-10-01 01:05:16  &  2.4 &  0.1  &  mono loose  &  7  &  66  &  9.17  &  -0.06  &  257  &  61.3 & 41.4 \\
GRB~161001A  &  &  55.2  &  1.1  &  mono loose  &  78  &  653  &  8.54  &  0.17  &  173  &  5.8 & 10.9 \\
GRB~170531B  & 2017-05-31 22:02:09  &  24.6  &  0.4  &  stereo loose  &  12  &  98  &  8.95  &  0.29  &  312  &  5.7& 3.4 \\
GRB~170531B  &    & 89.0   &  1.3  &  mono loose  &  62  &  556  &  8.68  &  -0.24  &  161  &  14.4 & 10.5 \\
GRB~171020A  &  2017-10-20 23:07:10  &  2.4  &  0.4  &  stereo loose  &  10  &  151  &  9.08  &  -1.67  &  380  &  1.2 & 1.6 \\
GRB~171020A  &    &  32.3  &  2.6  &  stereo loose  &  85  &  681  &  9.23  &  1.21  &  283  &  3.3 & 5.2 \\
GRB~180510A  &  2018-05-10 19:24:34  &  196.4  &  1.6  &  stereo loose  &  45  &  363  &  8.94  &  0.64  &  232  &  5.6 & 5.5 \\
GRB~180512A  &  2018-05-12 22:01:47  &  3.0 &  0.4  &  mono loose  &  21  &  149  &  8.82  &  0.9  &  257  &  12.0 & 17.0 \\
GRB~180512A  &   & 31.8  &  0.2  &  mono loose  &  2  &  47  &  8.0  &  -1.77  &  288  &  4.4 & 5.4 \\
GRB~180613A  &   2018-06-13 15:36:18  &  121.3 &  2.0  &  stereo loose  &  103  &  956  &  9.0  &  -0.29  &  191  &  3.0 & 4.8 \\
GRB~190627A  &  2019-06-27 11:18:31  &  390.8  &  0.4  &  stereo loose  &  14  &  91  &  8.84  &  1.03  &  419  &  7.7 & 5.6 \\
GRB~190821A  &  2019-08-21 17:10:03  &  46.8  &  1.7  &  stereo loose  &  46  &  450  &  9.28  &  -0.34  &  191  &  3.7 & 4.1\\
GRB~191004B  &  2019-10-04 21:33:41  &  30.1 &  0.4  &  stereo loose  &  5  &  55  &  9.99  &  -0.2  &  419  &  5.2 & 2.3\\
GRB~191004B  &   &   58.9  &  3.5  &  stereo loose  &  108  &  1069  &  9.42  &  -0.48  &  191  &  2.0 & 2.6 \\
\bottomrule
    \end{tabular}}
     \tablefoot{This table sumarises the results from H.E.S.S. observations of the \emph{loc} sample, mostly triggered by \Swift/BAT and \Fermi/LAT. The first column is the name of the GRB following the GCN convention. The second column, T$_0$, corresponds to the burst's onset time in UTC. The third and fourth columns provide the observation delay and acceptance-corrected exposure time, respectively. The fifth column provides the analysis configuration (see Sec.~\ref{sec:analysis} for details). The number of events detected in the ON and OFF regions, the acceptance-weighted exposure $\alpha_\mathrm{exp}$ and the statistical significances are given in the sixth to ninth columns. The energy threshold of the analysis is given in the tenth column, and the integral flux upper limits with $\alpha=-2.5$ and $\alpha=-5$ above the given energy threshold are provided in the last two columns. }

\end{table*}

\end{appendix}

\end{document}